%% file: main.tex
\RequirePackage{fix-cm}
\documentclass[smallextended]{svjour3}       
\smartqed  
\usepackage{graphicx}
\usepackage{algorithm}
\usepackage{algpseudocode}

\usepackage{amsmath}
\usepackage{amsfonts}
\usepackage{enumerate}
\usepackage[labelfont=bf]{caption}  
\usepackage{subfloat}

\usepackage[authoryear]{natbib}
\setcitestyle{aysep={}}
\usepackage{url}

\graphicspath{ {figures/} }

\newcommand{\R}{\mathfrak{R}}
\newcommand{\G}{\mathcal{G}}

 \journalname{ }

\begin{document}

\title{Gene tree reconciliation including transfers with replacement is hard and FPT 
}


\author{Damir Hasi\'{c}         \and
        Eric Tannier 
}


\institute{Damir Hasi\'{c} \at
              Department of Mathematics, Faculty of Science, University of Sarajevo, 71000 Sarajevo, Bosnia and Herzegovina \\
              \email{damir.hasic@gmail.com, d.hasic@pmf.unsa.ba}           
           \and
           Eric Tannier \at
              Inria Grenoble Rh\^one-Alpes, F-38334 Montbonnot, France\\
              Univ Lyon, Universit\'e Lyon 1, CNRS, Laboratoire de Biom\'etrie et Biologie \'Evolutive UMR5558, F-69622 Villeurbanne, France             
}

\date{ }

\maketitle

\begin{abstract}
    	\input{abstract}

    \end{abstract}
\input{introduction}

\input{definitions}

\input{np_hard}

\input{fpt_algorithm}

\input{equivalence_with_dated_spr}

\input{conclusion}
\input{figures}

\clearpage

\begin{acknowledgements}
	E.T. was supported by the French Agence Nationale de la Recherche (ANR) through grant no. ANR-10-BINF-01–01 ‘Ancestrome’.
	
	Conflict of Interest: The authors declare that they have no conflict of interest.
\end{acknowledgements}

\bibliographystyle{spbasic}      

\bibliography{bibliography} 
\end{document}

%% file: abstract.tex
Phylogenetic trees illustrate the evolutionary history of genes and species. In most cases, although genes evolve along with the species they belong to, a species tree and gene tree are not identical, because of evolutionary events at the gene level like duplication or transfer. These differences are handled by phylogenetic reconciliation, which formally is a mapping between gene tree nodes and species tree nodes and branches. We investigate models of reconciliation with a gene transfer that replaces existing gene, which is a biological important event but never included in reconciliation models. Also the problem is close to a dated version of the classical subtree prune and regraft (SPR) distance problem, where a pruned subtree has to be regrafted only on a branch closer to the root. We prove that the reconciliation problem including transfer and replacement is NP-hard, and that if speciations and transfers with replacement are the only allowed evolutionary events, then it is fixed-parameter tractable (FPT) with respect to the reconciliation's weight. We prove that the results extend to the dated SPR problem.

%% file: introduction.tex
\section{Introduction}

{\em Duplications} and {\em transfers} are events in evolution of genes, and one of the major reasons for discordance between species and gene trees. These differences are explained by {\em phylogenetic reconciliation} in \cite{Doyon2011}.

The main evolutionary event, investigated in this paper, is {\em gene transfer} ({\em horizontal gene transfer}, Figure \ref{fig:transfer}). It involves two (possibly ancient) species existing at the same moment. The species that provides a transfered gene is called a {\em donor species}, and the species that receives the gene is called  a {\em recipient species}. 

As phylogenetic analysis never includes the totality of living species, in particular ancient species which can be extinct or not sampled (see {\em transfer from the dead} in \cite{Szollosi2013}), the donor species is not assumed to belong to the species phylogeny, but is related to it through one of its ancestors. By extension, this ancestor is considered as the donor, which yields {\em diagonal transfers}, or {\em transfers to the future} (Figure \ref{fig:transfer} $(b-c)$).

The recipient species either receives a new gene copy, or replaces an existing one (Figure \ref{fig:transfer} $(d)$). The latter event is called {\em replacement transfer} or {\em transfer with replacement}. In \cite{Choi2012} replacement transfer is called {\em replacing horizontal gene transfer} (HGT). They found that the replacing HGT and the additive HGT affect differently gene functions in {\em Streptococcus}. In \cite{Rice2006} HGT in plastid genomes is studied and the evidence of transfer with replacement is found. HGT with replacement also occurred  in the evolution of eukaryotes (\cite{Keeling2008}).

In this article we explore the algorithmic aspects of transfers with replacements when time constraints are imposed on transfers ({\em i.e.} they should not be directed to the past).

\subsection{A review of previous results}

There are two usual ways of detecting transfers by comparing species trees and gene trees. One is reconciliation, the other is the computation of SPR scenarios. Our work lies at the intersection of the two. Indeed, reconciliations can consider time constraints but never include replacing transfers. On the other hand SPR scenarios are a good model for replacing transfers but never consider time constraints.

Time constraints result in ---fully or partially--- dated species trees.
For dated species tree, finding a reconciliation of minimum cost, in a model with gene transfer, is usually polynomial \citep{Merkle2010, Doyon2010, Bansal2012}.
With undated species tree, and partial time constraints it is usually NP-hard \citep{Hallett2001, Hallett2004, Bansal2012}, and can be fixed parameter tractable \citep{Hallett2001, Hallett2004}, or inapproximable \citep{Dasgupta2006}.
If a constraint on time consistency of reconciliation scenarios is relaxed, the problem becomes polynomial \citep{Hallett2004, Bansal2012}.

There are some results that go beyond finding one optimal solution. In \cite{Bansal2013} an algorithm that uniformly samples the space of optimal solutions is given, and it runs in polynomial time (per sample). In \cite{Chan2015} the space of all optimal solutions is also explored, and formula for the number of optimal solutions is given. Continuous and discrete frameworks for finding a minimum duplication-transfer-loss (DTL) reconciliation are equivalent \citep{Ranwez2016}. Only recently, DTL model is expanded by \cite{CHAN20171}, where incomplete lineage sorting is included, and FPT algorithm that returns a minimal reconciliation is given. Probabilistic models allow sampling of solutions in larger spaces, according to likelihood distributions \cite{Szollosi2013}.

Until now only reconciliations with transfers without replacements are investigated. However, note that duplications with replacement, {\em i.e.} {\em conversions}, were recently introduced by \cite{conversionPaper2017}. For a more detailed review on reconciliations see \cite{Szllosi2015}, \cite{Nakhleh2012}, and \cite{Doyon2011}. 

Transfers that replace existing genes are in close relation to the classical tree rearrangement operation {\em subtree prune and regraft} (SPR). For the definition of SPR refer to \cite{Song2006} (for the rooted trees) and \cite{Allen2001} (for the unrooted trees). This operation has never been integrated in reconciliation models but is used to detect transfers when it is the only allowed evolutionary event at the gene scale. In consequence these studies are limited to datasets where genes appear in at most one copy per species.

Computing SPR distance between two rooted binary phylogenetic trees on the same label set is NP-hard. The first proof \citep{Hein1995} has a mistake (see \cite{Allen2001}). Nevertheless, the results from \cite{Hein1995} can be used for {\em tree bisection and reconnection}. The correct proof is in \cite{Bordewich2005}. They used a modified approach from \cite{Hein1995}. The problem is also NP-hard for unrooted binary trees \citep{Hickey2008}. If we take the SPR distance as a parameter, then both rooted  \citep{Bordewich2005} and unrooted \citep{Whidden2015} versions are FPT.

There is an approximation algorithm of ratio 3 \citep{Hein1995}, as well as an ILP algorithm for calculating an exact rooted SPR distance \citep{Wu2009}. An exact rooted SPR distance is also determined by reducing it to CNF \citep{Bonet2009} and using existing SAT solvers.

A probabilistic model of gene transfer with replacement using (undated) SPR, where time consistent and {\em transfers to the past} are not distinguished, is given by \cite{Suchard2005}.
Model for host-parasite cophylogeny by \cite{huelsenbeck2000} could also be used to detect gene transfers between loci.     

Rooted SPR distance is equivalent to the number of trees in the {\em maximal agreement forest} (MAF)    \citep{Bordewich2005}, while in the unrooted version SPR distance is greater than or equal to MAF \citep{Allen2001}. 

A divide and conquer approach with MAF is used for computing an exact SPR distance in \cite{Linz2011}. A 2.5-approximation algorithm for the MAF problem on two rooted binary phylogenetic trees is presented in \cite{Shi2014}. In \cite{Chen2013} an FPT algorithm for rooted SPR, with complexity $O(2.344^ k\cdot n)$ is presented, which is an improvement compared to $O(2.42^ k\cdot n)$ \citep{Whidden2010}. Rooted SPR is investigated also for non binary trees \citep{Whidden2016}, and MAF for multiple trees \citep{Shi2014b}. For a more complete review see \cite{Shi2013} and \cite{Whidden2016}.     

Dated SPR, that is, SPR distance on a dated species tree, where only contemporaneous or transfers to the future are allowed, is mentioned in \cite{Song2006}, where it is investigated how many dated trees are one SPR operation away from a given dated tree. The complexity of the dated SPR distance computation is left open, and has an answer as a consequence of our results here.

\subsection{The contribution of this paper}

In this paper, we analyze the algorithmic complexity of finding a minimum reconciliation with replacing transfers, in the presence of a dated species tree. If speciations and replacement transfers are the only evolutionary events in the reconciliation, then finding an optimal dated SPR scenario is an equivalent problem. 

We define a model of reconciliation with gene transfer followed by a gene replacement, {\em i.e.} transfered gene replaces a gene that is already present in the recipient species (Figure \ref{fig:transfer} $(d)$). We will call this event {\em transfer with replacement}, and it is represented by a transfered gene and a loss (the gene that is replaced). We prove that finding a minimum reconciliation that includes transfer with replacement, as well as transfer, duplication and loss, is NP-hard. If speciation and transfer with replacement are the only allowed events, then it is fixed parameter tractable with respect to the output size, and it is easily reducible to dated SPR problem. Therefore dated SPR is also NP-hard and FPT. Note that the hardness of dated SPR is not easily deduced from the hardness of general SPR because all the known proofs make an extensive usage of the possibility of time inconsistent SPRs.

We prove NP-hardness by a reduction from \textproc{Max 2-Sat}. The gadgets for the variables and clauses are constructed, and used to assemble a reconciliation that we call a {\em proper reconciliation}. Hence gadgets are sub-reconciliations of a proper reconciliation. Next, we state an obvious claim that relates an optimal \textproc{Max 2-Sat} solution with an optimal proper reconciliation. Then we prove that any optimal reconciliation can be transformed (in polynomial time) into a proper reconciliation of the same weight (therefore optimal).

In order to prove parametrized tractability, we introduce a {\em normalized reconciliation}. Intuitively, this reconciliation can be obtained from any reconciliation by raising nodes of $G$  as much as possible, that is, mapping them to species tree nodes and edges closer to the root, without affecting transfers, hence keeping the weight of a reconciliation. Then we give a branch and bound algorithm that returns an optimal reconciliation that is also normalized. Thanks to the normalization, we can have at most three cases in the branching algorithm, and every branching produces at least one transfer, so the depth of any branching procedure is at most $k$, {\em i.e.} $3^k\cdot n$ is an approximate complexity of the algorithm.

%% file: definitions.tex
\section{Definitions}

A {\em phylogenetic tree} here is a rooted tree $T$, such that the root vertex $root(T)$ has degree 1, and the incident edge is called {\em the root edge}, or degree of $root(T)$ is 2 and there is no root edge. With $L(T)$ is denoted the set of all leaves of the tree $T$. Trees are considered binary, meaning that the nodes have at most two children. We say that they are fully binary when all internal nodes have exactly two children. If $x$ is a node/edge in a rooted tree $T$, then $p_T(x)=p(x)$ denotes its parent, $x_l,x_r$ denotes its (left and right) children.

If $x$ is an ancestor of $y$ in a rooted tree $T$, {\em i.e.} if $x$ is in the path from $y$ to $root(T)$, then we write $y\le_T x$ or $y\le x$, defining a partial order on the nodes. Also, let $e_1=(p(s_1), s_1),e_2=(p(s_2), s_2)\in E(T)$ and $p(s_1)\le s_2$, then we can write $s_1<e_1<p(s_1)\le s_2<e_2<p(s_2)$. With this, we define a partial order on the set $V(T)\cup E(T)$. 
If $x$ is a node/edge in $T$, then $T(x)$ is the maximal rooted subtree with root node/edge $x$, and if $A\subseteq L(T)$, then $T(A)$ is the subtree of $T$ with a root vertex of degree 2, and $L(T(A))=A$. 

The next definition extends the partial order on the set $V(T)\cup E(T)$ to the total order by introducing the {\em date function}. Intuitively, to every node and edge from $T$ a date ({\em i.e.} a point in the past) is assigned. This derives from the fact that phylogenetic trees and reconciliations represent evolutionary events that happened at some point in the past.   

\begin{definition}[Date function. Dated tree]
Let $T$ be a rooted tree and  $\tau:V(T)\cup E(T)\rightarrow [0,+\infty)$ such that $\tau(L(T))=\{0\}$, $x_1,x_1\in V(T)\cup E(T)$, $x_1<x_2$ $\implies \tau(x_1)<\tau(x_2)$. Function $\tau=\tau_T$ is a \emph{date function} on the tree $T$, and $T$ is a {\em dated tree}.
\end{definition}

Note that the edges of $T$ are assigned a date. Although it might seem more natural to assign an interval to an edge, here it is more convenient to assign a point ({\em i.e.} a date).   

By a species tree $S$ we mean a dated, fully binary tree with function $\tau_S=\tau$, and $\tau(s_1)\ne \tau(s_2), \forall s_1,s_2\in V(S)\backslash L(S), s_1\ne s_2$.
{\em Subdividing} an edge means that a vertex is added to the edge. Formally, edge $e=(x,y)$ is subdivided if a node $z$ is added to the graph along with edges $(x,z), (y,z)$, and the edge $e$ is removed. 

\begin{definition}[Subdivision of a species tree]\label{def:treeSubdivision}
	Let $S'$ be a tree obtained from $S$ by subdividing some edges, and $\forall e=(p(s), s)\in E(S)$, and $\forall s_1\in V(S)$ for which $\tau(s)<\tau(s_1)<\tau(p(s))$, $\exists s'\in V(S')\backslash V(S)$, $\tau(s_1)= \tau(s')$, and $s<s'<p(s)$. Tree $S'$ with these properties, and with the minimum number of nodes is called the {\em subdivision} of the species tree $S$.
\end{definition}

Note that the node $s'$ from Definition \ref{def:treeSubdivision} is obtained by subdividing some edge, and $deg(s')=2$. Subdivision of a species tree is unique (Figure \ref{fig:subdivision}). Also, $L(S')=L(S)$ and $root(S')=root(S).$ If $e\in E(S')$, then $v_e$ denotes the maximum element from the set $\{x\in V(S)\mid x<e\}$. We assume that $\tau(V(S')\cup E(S'))=\{0,1,\ldots,2n\}$ (see Figure \ref{fig:subdivision}), where $n$ is the number of the extant species in $S$, $\tau(L(S))=\{0\}$, $\tau(root(S))=2n$. Therefore if $x\in V(S')\cup E(S')$, then $\tau(p_{S'}(x))=\tau(x)+1$. 

We now define gene tree species tree reconciliations. Note $G$ a gene tree, which is a fully binary tree which comes with a mapping $\phi:L(G) \rightarrow L(S)$ that indicates the species in which genes are found in the data.

\begin{definition}[Extension]
	A tree $T'$ is said to be an {\em extension} of a tree $T$ if $T$ can be obtained from $T'$ by pruning some subtrees and suppressing nodes of degree 2.    
\end{definition}

The next definition is fundamental for the notion of phylogenetic reconciliation. Function $\rho$ indicates positions of  genes inside the species tree.  

\begin{definition}[Semi-reconciliation]
	Let $G'$ be an extension of a gene tree $G$, and $S$ is a species tree. Let $\phi:L(G)\rightarrow L(S)$, and $\rho:V(G')\rightarrow V(S)\cup E(S')$ such that $\rho/L(G) = \phi$ and $\rho(root(G'))=\rho(root(G))=root(S)$. If $x,y\in V(G')$, $x<y$, and $\rho(x)$ and $\rho(y)$ are comparable in $S$, then $\rho(x)\le \rho(y)$.  The 6-tuple $\R=(G,G',S,\phi,\rho, \tau)$ is called {\em semi-reconciliation}.
\end{definition}

Note that the nodes from $G'$ are not mapped into $V(S')\backslash V(S)$. If $\rho(x)=e'\in E(S')$ and $e'$ is a part of $e\in E(S)$, then we will write $\rho(x)\in e$. 

The next definition introduces the notion of subtree of $G'$ that is not in $G$.

\begin{definition}[Lost subtree]
	Let $\R$ be a reconciliation. A maximal subtree $T$ of $G'$ such that $V(T)\cap V(G)=\emptyset$ is called a {\em lost subtree}.
\end{definition}

\begin{definition}[Sub-branch]
	If $G'$ is an extension of $G$, $(x_1,x_2)\in E(G)$, $(x'_1,x'_2)\in E(G')$ and $x_2\le x'_2\le x'_1 \le x_1$, then we say that $(x'_1,x'_2)$ is {\em a sub-branch} of $(x_1,x_2)$. Similarly, we can define sub-branch for $S$ and $S'$.   
\end{definition}

Semi-reconciliation is a reconciliation without established evolutionary events. The next definitions introduce these events.

\begin{definition}[Speciation]
	Let $\R$ be a semi-reconciliation, $x\in V(G')$, $x'_l,x'_r$ are the children of $x$ in $G'$. Let  $\rho(x)\in V(S)\backslash L(S)$ and $\rho(x)_l \le \rho(x'_l) < \rho(x)$, $\rho(x)_r \le \rho(x'_r) < \rho(x)$. Then $x$ is called a {\em speciation}. The set of all speciations is denoted by $\Sigma(\R)$ or $\Sigma$.   
\end{definition}

\begin{definition}[Duplication]
	Let $\R$ be a semi-reconciliation, $x\in V(G')$, $x'_l,x'_r$ are the children of $x$ in $G'$. Let $\rho(x)=e\in E(S')$, $v_e\le \rho(x'_l)$, and $v_e\le \rho(x'_r)$. Then $x$ is called a {\em duplication}. The set of all duplications is denoted by $\Delta(\R)$ or $\Delta$. 
\end{definition}

From now on, we will assume $\tau(x)=\tau(\rho(x))$, for all $x\in V(G')$.

\begin{definition}[Transfer]
	Let $\R$ be a semi-reconciliation, $x\in V(G')$, $x'_l,x'_r$ are the children of $x$ in $G'$, $\rho(x)=e\in E(S')$ and for one of the $\rho(x'_l),\rho(x'_r)$ (say $\rho(x'_l)$) holds $v_{e}\le \rho(x'_l)$ and for the other one (i.e. $\rho(x'_r)$) $\rho(x'_r)=e'\in E(S')$, $\tau(e')\le \tau(e)$, $deg(x'_r)=2$, and $v_{e'}\le \rho(x''_r)$, where $x''_r$ is the only child of $x'_r$ in $G'$. Then $x$ is called a {\em transfer parent}, $x'_r$ is  a \emph{transfer child}, and the edge $e=(x,x'_r)\in E(G')$ is a \emph{transfer}. If $\tau(x'_r)=\tau(x)$, the transfer is {\em horizontal transfer}, and if $\tau(x'_r) < \tau(x)$, the transfer is {\em diagonal transfer} or {\em transfer to the future}. The set of all transfers is denoted by $\Theta(\R)$ or $\Theta$. 
\end{definition}

\begin{definition}[Loss]
	Let $\R$ be a semi-reconciliation, and $x\in L(G')\backslash L(G)$. Then $x$ is called a \emph{loss}. The set of all losses is denoted by $\Lambda(\R)$ or $\Lambda$.
\end{definition}

The next two events that we are going to define are created by pairing some of the previously defined events with a loss.

\begin{definition}[Replacement transfer]
	Let $(G,G',S,\phi,\rho, \tau)$ be a semi-reconciliation, $\delta_T: \Theta \rightarrow \Lambda $ is an injective partial function such that $\rho(x_2)=\rho(\delta_T(e))$ for all $e\in \delta^{-1}_T(\Lambda)$, where $e=(x_1,x_2) \in \Theta$. If $e\in \delta^{-1}_T(\Lambda)$, then $e$ is called a {\em replacement transfer} or {\em transfer with replacement}, and $l=\delta_T(e)$ is its associate loss. The set of all replacement transfers is denoted by $\Theta'$ and the set of all associate losses by $\Lambda'_T$.  
\end{definition}

In the previous definition, mapping $\delta_T$ pairs transfer $e$ (or we can say a transfer child $x_2$) with the loss $l$ (see Figure \ref{fig:transfer}). In this way, we get that gene $x_2$ is replacing gene $l$, hence the name {\em transfer with replacement}. Requirement $\rho(x_2)=\rho(l)$ is necessary if $x_2$ replaces $l$.

Conversion is to duplication what replacement transfer is to transfer. 

\begin{definition}[Conversion]
	Let $(G,G',S,\phi,\rho, \tau)$ be a semi-reconciliation, $\delta_D: \Delta \rightarrow \Lambda $ is an injective partial function such that $\rho(x)=\rho(\delta_D(x))$ for all $x\in \delta^{-1}_D(\Lambda)$. If $x\in \delta^{-1}_D(\Lambda)$, then $x$ is called a {\em conversion}, and $\delta_D(x)$ is its associate loss. The set of all conversions is denoted by $\Delta'$ and the set of associate losses by $\Lambda'_D$. 
\end{definition}

The elements of $\Lambda'=\Lambda'_T\cup\Lambda'_D$ are called \emph{free losses}. The set of all (evolutionary) events is $\{S, D, T, L, C, T_R\}$.

\begin{definition}[Reconciliation]
	Let $(G,G',S,\phi,\rho, \tau)$ be a semi-reconciliation, and $A\subseteq \{D, T, L, C, T_R\}$. To every node from $V(G')\backslash L(G)$  some event from $A\cup\{S\}$ is attached. Then $\R=(G,G',S,\phi,\rho,\tau,\delta_T,\delta_D,A)$ is called {\em A reconciliation}.
\end{definition}

If transfers with replacement or conversions are not included in a reconciliation, then $\delta^{-1}_T(\Lambda)=\emptyset$, or $\delta^{-1}_D(\Lambda)=\emptyset$.
Note that if $x\in V(G')$ and $deg(x)=2$, then $(p_{G'}(x), x)\in \Theta(\R)$.

Speciations, duplications, transfers, losses, conversions, and transfers with replacement are called {\em evolutionary events}. A reconciliation can allow only some of these events. For example, if a reconciliation $\R$ allows speciations, duplications and losses, we will call it {\em DL reconciliation}. If $\R$ also allows transfers, we call it {\em DTL reconciliation}. Speciations are assumed to be allowed in every reconciliation, so they are not emphasized in the type of a reconciliation. If transfers are not allowed in a reconciliation, then the date function is not necessary, and can be disregarded. Note that if $A\subseteq B$, then any $A$ reconciliation is also a $B$ reconciliation. If conversions or transfers with replacement are included in a reconciliation, then we assume that free losses are allowed. Therefore $T_R$ reconciliation allows speciations, replacement transfers, and free losses, while $T_RL$ reconciliation additionally allows non-free losses. 

Not every semi-reconciliation can produce a reconciliation. For example, if a node from $G'$ is mapped under its LCA ({\em Last Common Ancestor} - see \cite{Goodman1979}, \cite{Chauve2009}) position, then the transfers must be allowed as an event in order to obtain a reconciliation.

\begin{definition}[Weighted reconciliation]
	Let $\R$ be an $A$ reconciliation, and $A=\{a_1,\ldots,a_k\}$. If $c_i\ge 0$ are associated with the events $a_i$ $(i=1,\ldots,k)$, then $\omega(\R)=\sum c_i\cdot |a_i|$  is called {\em the weight or cost of} $\R$, where $|a_i|$ denotes the number of nodes in $G'$ that are associated with the event $a_i$, for $i=1,\ldots,k$. 
\end{definition}

We see that speciations do not affect the weight of a reconciliation, thus take that their weight is 0. In this paper free losses (losses assigned to a conversion or replacement transfer) have weight 0. Other events, included in a reconciliation, have weight 1.

\begin{definition}[\textproc{Minimum A Reconciliation} problem]
	Let $G$ and $S$ be gene and species trees. The problem of finding an $A$ reconciliation of minimum weight is called {\em \textproc{Minimum A Reconciliation}}. 
\end{definition}

The next definition introduces the weight of a subtree of $G'$. This is necessary because we estimate the weight of a reconciliation by decomposing $G'$ into subtrees and evaluating the weight of every subtree. 

\begin{definition}[The weight of a subtree]
	Let $\R$ be a reconciliation and $T$ is a subtree of $G'$. By $\omega_{\R}(T)$ or $\omega(T)$ is denoted the sum of weights of all events assigned to the nodes and edges of $T$.
\end{definition}

%% file: np_hard.tex
\section{Finding an optimal $\mathbf{DTLCT_R}$ reconciliation is NP-hard}

In this section, and the rest of the paper, we assume that all events are of weight 1, except speciations and free losses, which are of weight 0. We prove that finding a minimum reconciliation that includes transfers with replacement is NP-hard. We first prove the NP-hardness of the problem of finding a minimum reconciliation that includes all events (duplication, transfer, loss, conversion, transfer with replacement).

We will use a reduction from \textproc{Max 2-Sat}.

\noindent\textproc{Max 2-Sat:}\\
\textbf{Input:} $F=C_1 \land C_2 \land \ldots \land C_m$; $C_j=x'_{j_1} \lor x'_{j_2}$, $j=1,\ldots,m$;
$K\le m$.\\
\textbf{Output:} Is there a truth assignment for logical variables $x_1,\ldots,x_n$  such that there are at least $K$ true clauses. 

This problem is NP-hard \citep{GAREY1976237,Garey:1979:CIG:578533}, solvable in polynomial time if $K=m$ \citep{Even76,Garey:1979:CIG:578533}. It remains NP-hard even if every variable appears in at most three clauses \citep{RAMAN19981}. We assume that every variable appears in exactly three clauses, and both positive and negative literals are present. We also assume the optimization version of this problem that asks for the minimum number of false clauses.

Formula $F$ is called a logical expression/formula. Variables $x_1,\ldots, x_n$ are (logical) variables. If $x$ is a variable, then $x$ is called a {\em positive literal}, and $\neg x$ is a {\em negative literal}. To a variable $x_i$ literals $x^1_i,x^2_i,x^3_i$ are assigned. We can assume that $x^1_i$ and $x^2_i$ have the same logical value, which is different from the logical value of $x^3_i$. 
Variables can be {\em true} or {\em false}. Literal is {\em true} if it is positive and the variable is true, or if it is negative and the variable is false. Similarly, literal is {\em false} if it is positive and the variable is false, or if it is negative and the variable is true.

\subsection{Variable and clause gadgets}

In order to construct a polynomial reduction from \textproc{Max 2-Sat} to \textproc{Optimal $DTLCT_R$ Reconciliation}, suppose we have a logical formula $F$ of \textproc{Max 2-Sat}, with $n$ variables and $m$ clauses, such that each variable appears exactly three times as a literal, and both positive and negative literals are present. We will construct a species tree, a gene tree, and a function $\phi$ mapping the gene tree leaves to the species tree leaves, an instance of the reconciliation problem.

First, we  introduce the \emph{border line} that corresponds to some date, depicted by horizontal  dashed line in Figures \ref{fig:variable_gadget}, \ref{fig:variable_gadgetb}, \ref{fig:clause_gadgeta}, \ref{fig:clause_gadgetb}, \ref{fig:whole_reconciliation},  \ref{fig:example_whole_reconciliation}.
Some nodes of the constructed gene tree will be assigned to literals of $x_i$ ($i=1,\ldots,n$), and in an optimal reconciliation, their mapping above or under this border line will decide if the literals are true or false. In consequence, the positive and negative version of a same variable must be mapped on the opposite sides of the border line in reconciliations.

For each variable and each clause we define a piece of a gene tree and a piece of a species tree with appropriate function $\phi$. The gadget for a variable $x_i$ is illustrated in Figure \ref{fig:variable_gadget}. The species subtree $S_{x_i}$ consists in 28 leaves named $A_1^i,\dots,A_{28}^i$, organized in two subtrees. Seven cherry trees are under the border line on each part, and then linked by two combs, one fully above and one fully under the border line.
The gene subtree $G_{x_i}$ is also organized in two subtrees, each consisting in 7 cherry trees linked by a comb. One of the subtrees is identified as the "positive literal subtree" and the other as the "negative literal subtree". The function $\phi$ mapping the leaves of the gene tree and the leaves of the species tree is such that 
$\phi(r_{i,k})=A^i_{2k-1}$ ($k\in \{1,\ldots,7\}$), $\phi(r_{i,k})=A^i_{2k}$ ($k\in \{8,\ldots,14\}$), $\phi(l_{i,k})=A^i_{29-2k}$ ($k\in \{1,\ldots,7\}$), $\phi(l_{i,k})=A^i_{30-2k}$ ($k\in \{8,\ldots,14\}$), where $(r_{i,k},l_{i,k})$ is the $k$th cherry of the gene tree $G_{x_i}$ (i.e. $r_{i,k},l_{i,k}$ are the children of $b^k_i$).

Then both trees $G$ and $S$ are anchored by an outgroup comb of size $P(n)$, a polynomial with sufficient size, with respective leaf sets $a^1_i,\dots,a^{P(n)}_i$ and $A_{i,1},\dots,A_{i,P(n)}$, and $\phi(a^k_i)=A_{i,1}$ as illustrated by Figure \ref{fig:variable_gadgetb}. 

Figures \ref{fig:clause_gadgeta} and \ref{fig:clause_gadgetb} illustrate the gadget for a clause $C_j$. The species subtree $S_{C_j}$ is a fully balanced binary tree with 8 leaves, noted $B_1^j,\dots,B_8^j$. The internal nodes of the subtree leading to $B_1^j,\dots,B_4^j$ are all above the border line, while the internal nodes of the subtree leading to $B_5^j,\dots,B_8^j$ are all under the border line. To each literal from the clause corresponds a fully balanced gene tree with four leaves, respectively mapping by function $\phi$ to $((B^j_1,B^j_7),(B^j_2,B^j_6))$ and $((B^j_3,B^j_5),(B^j_4,B^j_8))$ (which is an arbitrary way of mapping each cherry to the two different species subtrees). The internal nodes of the two gene subtrees are respectively labeled $r_{j1}$, $r_{j1}^0$, $r_{j1}^1$, and $r_{j2}$, $r_{j2}^0$, $r_{j2}^1$. The forest of these two gene subtrees is noted $F_j$.

Clauses and variables gadgets are linked to form the full trees $G$ and $S$. First, each gene subtree in a clause $C_j$, representing a literal of a variable $x_i$, is linked by its root $r_{j_1}$ or $r_{j_2}$ to $G_{x_i}$ (i.e. to $x^1_i$, $x^2_i$, or $x^3_i$), in the middle of the comb of the appropriate subtree (positive literal subtree if the literal is positive, or conversely). Second, the species subtrees and the gene subtrees are linked by a comb containing all variables and clauses in the order $x_1,\dots,x_n,C_1,\dots,C_m$ as described by Figure \ref{fig:whole_reconciliation}.

\subsection{Proper reconciliation}

Now that we have constructed an instance for the reconciliation problem from a logical formula, we need to be able to translate a reconciliation into an assignment of the variables. This is possible for a type of reconciliation named {\em proper}. Proper reconciliations are illustrated in Figure \ref{fig:whole_reconciliation}. 

\begin{definition}[Proper reconciliation]\label{def:proper}
We call a reconciliation $\R=(G,G',$ $S,\phi,\rho,\tau,\delta_T,\delta_D,\{T_R\})$, where $G$ and $S$ are constructed from a logical formula, a {\em proper} reconciliation if
\begin{itemize}
\item all transfers are horizontal;
\item in variable gadgets, the gene tree vertices in the anchor comb are mapped by $\rho$ to the species tree vertices in the anchor comb, that is, $\rho(c^0_i)=s^0_{x_i}$, $\rho(d^k_i)=D^i_k$ (for all $k\in \{1,\ldots, P(n)\}$), $\rho(d_i)=s^1_{x_i}$;
\item in variable gadgets, the two gene tree comb internal vertices ($c_i^k$ in the figure) are mapped to the two species tree combs (vertices $C'_{i,k}$ in the Figure), in one of the two possible combinations (the two gene tree combs may map to either species tree combs). 
\item in clause gadgets, the mapping $\phi$ corresponds to one of the cases drawn on Figures \ref{fig:clause_gadgeta} and \ref{fig:clause_gadgetb}, that is:
	\begin{itemize}
		\item $\rho(r_{j_1})=B^j_{1,2}$, $\rho(r_{j_2})=B^j_{5,6,7,8}$, $\rho(r^0_{j_1})\in (B^j_{1,2}, B^j_1)$,  $\rho(r^1_{j_1})\in (B^j_{1,2}, B^j_2)$, $\rho(r^0_{j_2})\in (B^j_{5,6}, B^j_5)$, $\rho(r^1_{j_2})\in (B^j_{7,8}, B^j_8)$;
		
		\item  $\rho(r_{j_1})=B^j_{5,6,7,8}$, $\rho(r_{j_2})=B^j_{3,4}$, $\rho(r^0_{j_1})\in (B^j_{5,6}, B^j_6)$,  $\rho(r^1_{j_1})\in (B^j_{7,8}, B^j_7)$, $\rho(r^0_{j_2})\in (B^j_{3,4}, B^j_3)$, $\rho(r^1_{j_2})\in (B^j_{3,4}, B^j_4)$;
		
		\item $\rho(r_{j_1})=B^j_{1,2}$, $\rho(r_{j_2})=B^j_{3,4}$, $\rho(r^0_{j_1})\in (B^j_{1,2}, B^j_1)$,  $\rho(r^1_{j_1})\in (B^j_{1,2}, B^j_2)$, $\rho(r^0_{j_2})\in (B^j_{3,4}, B^j_3)$, $\rho(r^1_{j_2})\in (B^j_{3,4}, B^j_4)$;
		
		\item $\rho(r_{j_1})=B^j_{5,6,7,8}$, $\rho(r_{j_2})\in (B^j_{3,4},B^j_3)$, $\rho(r^0_{j_1})\in (B^j_{5,6}, B^j_6)$,  $\rho(r^1_{j_1})\in (B^j_{7,8}, B^j_7)$, $\rho(r^0_{j_2})\in (B^j_{3,4}, B^j_3)$, $\rho(r^1_{j_2})\in (B^j_{3,4}, B^j_4)$;	\end{itemize} 
\end{itemize}
\end{definition}
 
Note that a proper reconciliation is a $T_R$ reconciliation, {\em i.e.} the only events are speciations, replacement transfers, and free losses. Hence the weight of a proper reconciliation is the number of transfers.

Let $F$ be a logical formula and $G$, $S$ are gene and species tree assigned to $F$, as previously described. There is an obvious relation between value assignment to logical variables and a proper reconciliation between $G$ and $S$. The next lemma and its proof describes and quantifies this relation.

\begin{lemma}\label{formula->properReconciliation}
	Let  $F=C_1 \land C_2 \land \ldots \land C_m$ be a logical formula, and $G,S$  are gene and species trees assigned to $F$. Let $\R$ be a proper reconciliation between $G$ and $S$. There is an assignment of the logical variables which satisfies exactly $17n+5m+\omega(\R)$ clauses.
\end{lemma}
\begin{proof}
The assignment is constructed from the proper reconciliation according to the positions of the corresponding vertices above or under the border line. The definition of proper reconciliation ensures that two opposite literals are always on the opposite side of the border line.
	Every variable gadget has 17 transfers (counting the ones incident with $x^1_i,x^2_i,x^3_i$), and the total number of transfers generated by these gadgets is $17n$.
    Clause gadget has 5 or 4 transfers (not counting incoming transfers, because they are already counted at the variable gadgets), depending if the clause's literals are both false or not. We have $f$ clause gadgets with 5 transfers corresponding to unsatisfied clauses. Hence the number of transfers generated in the clause gadgets is $4(m-f)+5f=4m+f$. This yields $\omega(\R)=17n+4m+f$, so there are $17n+5m+\omega(\R)$ satisfied clauses.
    \qed
\end{proof}

We see that if we minimize the cost of a proper reconciliation, we also minimize the number of false clauses in the logical formula. As an immediate consequence of Lemma \ref{formula->properReconciliation}, we have the next lemma.

\begin{lemma}
To a proper optimal reconciliation corresponds an optimal logical formula. 
\end{lemma}

In order to prove NP-hardness, we need to show that there is an optimal, proper reconciliation, which can be easily (in polynomial time) obtained from an arbitrary optimal ($DTLCT_R$) reconciliation. 

\subsection{Optimal proper reconciliation}

In this section we describe how to construct a proper optimal reconciliation, given  an optimal reconciliation.  

Let $\R$ be an arbitrary reconciliation. If $\R$ is a proper reconciliation, then $\omega(G_{x_i})=17$ (here we also count three transfers incident with $x^1_i,x^2_i,x^3_i$), and $\omega(F_j)\in \{4,5\}$ for all variable and clause gadgets.

\begin{lemma}\label{lem:nodePositions}
	Let $\R$ be a reconciliation between $G$ and $S$, where $G$ and $S$ are gene and species trees constructed from a logical formula, and $i\in \{1,\ldots, n\}$, $j \in \{1,\ldots, m\}$. We have:
	\begin{enumerate}[(a)]
		\item $\omega(F_j)\ge 4$;
		\item if both  $r_{j_1}$ and  $r_{j_2}$ are under the border line, then $\omega(F_j) \ge 5$;
		\item $\omega(G_{x_i})\ge 17$;
		\item if $x^1_i$ or $x^2_i$ is on the same side of the border line as $x^3_i$, then $\omega(G_{x_i})\ge 19$.
	\end{enumerate}  
\end{lemma}
\begin{proof}

    To prove $(a)$ and $(b)$, we identify two cases, according to the position of $r_{j_1}$ and  $r_{j_2}$.	
	
	Case 1. Node $r_{j_1}$ or $r_{j_2}$ is above the border line. In $\mathcal{G}_{C_j}$ each of $r^0_{j_1}, r^0_{j_2}, r^1_{j_1}, r^1_{j_2}$ is incident with exactly one transfer. In order to obtain $\omega(F_j)<4$, we need to achieve that some of the nodes $r^0_{j_1}, r^0_{j_2}, r^1_{j_1}, r^1_{j_2}$ is neither duplication nor incident with a transfer. The only way to achieve this is to place some of them in $s^0_{C_j}=root(S_{C_j})$. Let us take $\rho(r^0_{j_1}) = s^0_{C_j}$. Then $\rho(r_{j_1}) > s^0_{C_j}$, or  $\rho(r_{j_1})$ and $s^0_{C_j}$ are incomparable. If $\rho(r^1_{j_1}) < s^0_{C_j}$, then $(r_{j_1},r^1_{j_1})$ is a transfer, and the weight of $F_j$ is not decreased. If $\rho(r^1_{j_1}) = s^0_{C_j}$, then $r_{j_1}$ is a duplication, or one of the edges $(r_{j_1}, r^0_{j_1})$ and $(r_{j_1}, r^1_{j_1})$ contains a transfer. In this way we eliminate two transfers (that were incident with $r^0_{j_1}$ and $r^1_{j_1}$), and obtain one transfer or duplication. But we generate at least one non-free loss in $S_{C_j}$. Similar considerations apply to the other nodes of $F_j$. Hence we cannot obtain $\omega(F_j)<4$.  
	
	Case 2. Both nodes $r_{j_1}$ and $r_{j_2}$ are under the border line. None of the nodes $r^0_{j_1}, r^1_{j_1}, r^0_{j_2}, r^1_{j_2}$ can be placed at $s^0_{C_j}$, therefore every one of them is incident with at least one transfer.  If we wish to eliminate transfers starting at $r_{j_1}$ or $r_{j_2}$, then we need to place them both in $lca(B_5,B_6,B_7,B_8)$, i.e. in the minimal node in $S_{C_j}$ that is ancestor of $B_5$, $B_6$, $B_7$, and $B_8$ (in Figure \ref{fig:clause_gadgetb} $(d)$ the node that is placed in $lca(B_5,B_6,B_7,B_8)$ is $r_{j_1}$, i.e. $\rho(r_{j_1})=lca(B_5,B_6,B_7,B_8)$). In this case we increase the number of non-free losses. Whichever placement we choose, we have $\omega(F_j)\ge 5$. 
	
	$(c)$ The proof is similar in spirit to the proof of $(a)$. Observe Figures \ref{fig:variable_gadget} and \ref{fig:variable_gadgetb}. First, note that moving nodes (i.e. speciations) that belong to the part of variable gadget called anchor (i.e. nodes $d^1_i,\ldots,d^{P(n)}_i$), in order to position some nodes from $G_{x_i}$ would produce extra transfers. Also, moving all $P(n)$ nodes would create a reconciliation more expensive than any proper reconciliation. Hence we can assume that anchoring nodes $d^1_i,\ldots,d^{P(n)}_i$ are not moved. Because if this, when we move nodes of $G_{x_i}$ out of $S_{x_i}$, we cannot raise them above $D^i_1=p(s^0_{x_i})$, hence transfers are created.   
	
	There are 14 transfers incident with $b^s_i$ $(s=1,\ldots, 14)$. We can achieve that no transfer or duplication is incident with $b^s_i$ only if $\rho(b^s_i)=s^0_{x_i}$. Then a parent of $b^s_i$ (i.e. $c^{s-1}_i$), as well as $c^0_i$,  becomes a duplication, or is incident with a transfer, and two or more (depending on which $b^s_i$ is moved) non-free losses are created. Therefore taking $\rho(b^s_i)=s^0_{x_i}$, for some values of $s$, does not decrease $\omega(G_{x_i})$.

	Observe transfers incident with $x^1_i, x^2_i$, and eliminate them by moving nodes $x^1_i, x^2_i$ to an appropriate node of $S$, and assume that $(x^2_i, x^1_i)$ is not a transfer. By eliminating these transfers, we obtain at least two new transfers (at edges $(c^4_i, x^2_i), (x^1_i, c^3_i)$, or at some other edges). Similar considerations apply for $x^3_i$. Therefore, in this case too we cannot decrease the number of transfers.    
	
	Can we have less than 17 transfers if take $\rho(b^s_i)=s^0_{x_i}$, for some values of $s$, and eliminate transfers incident with $x^1_i, x^2_i$ ? Let us take $\rho(b^7_i)=s^0_{x_i}$. Then we need to move nodes $c^6_i$ and $c^0_i$, which generates at least two new transfers or duplications, and new non-free losses. Also, moving nodes $x^1_i,x^2_i$ will generate at least one transfer, different from the previous two newly generated. Hence we have at least three new events, and we cannot obtain less than 17 events (transfers, duplications, non-free losses).   
	
	$(d)$ Let us take that $x^1_i$ and $x^3_i$ are under the border line. Then at least three of the nodes $c^1_i,\ldots, c^{12}_i$ are not on the gadgets positions. Some of these nodes are $c^1_i,c^2_i,c^3_i$, because they are descendants of $x^1_i$ in $G$.  The paths $(c^1_i,b^2_i, A^3_i)$, $(c^2_i,b^3_i, A^5_i)$, $(c^3_i,b^4_i, A^7_i)$ generate extra three transfers. An extra transfer is created on the edge $(x^2_i,x^1_i)$, or at some other edge, if we move $x^2_i$ together with $x^1_i$. Even if we assume that, by moving $x^1_i$ (and $x^2_i$), we eliminate two transfers that were incident with them, we gain 4 more. Hence $\omega(G_{x_i})\ge 19$. \qed 
\end{proof}

The proof of next theorem describes a polynomial algorithm that transforms an optimal reconciliation $\R$ into a reconciliation $\R'$ that is both optimal and proper.  

\begin{theorem}\label{th:optimalProper}
	Let $G$ and $S$ be a gene and species tree. There is an optimal reconciliation that is proper. 
\end{theorem}
\begin{proof}
  Let $\R$ be an optimal reconciliation. We use $\R$ to construct $\R'$ that is both optimal and proper.
  
  Move the vertices of $G_{x_i}$ and position them in $S_{x_i}$ to obtain $\rho(c^0_i)=s^0_{x_i}$, $\rho(d^k_i)=D^i_k$ (for all $k\in \{1,\ldots, P(n)\}$), $\rho(d_i)=s^1_{x_i}$ $(i=1,\ldots, n)$. If $x^1_i$ and $x^2_i$ were not on the same side of the border line as $x^3_i$ (in $\R$), then they remain on the same side in $\R'$ as in $\R$. If $x^1_i$ or $x^2_i$ was on the same side as $x^3_i$ (in $\R$), then place $x^1_i, x^2_i$ above, and $x^3_i$ under the border line (in $\R'$). 
  
  Next, move the vertices of $F_j$ (in $\R$) and position them in $S_{C_j}$ to obtain the conditions of Definition \ref{def:proper}. Nodes $r_{j_1}$ and $r_{j_2}$ are positioned on the same side of the border line as $x'_{j_1}$ and $x'_{j_2}$, respectively. A reconciliation, obtained in this way, denote by $\R'$. It is obvious that $\R'$ is a proper reconciliation (by the construction). Let us prove that it is an optimal reconciliation. 
  
  We have $\omega_{\R}(G_{x_i})\ge 17 = \omega_{\R'}(G_{x_i})$, $\omega_{\R}(F_j)\ge 4$, and $\omega_{\R'}(F_j)\in \{4, 5\}$ (Lemma \ref{lem:nodePositions}). 
  
  Let $i\in \{1,\ldots,n\}$, $x^1_i,x^2_i,x^3_i$ are connected with $r_{a_1}\in V(F_a), r_{b_1}\in V(F_b),$ $r_{c_1}\in V(F_c)$ via transfers, and $\Omega_{\R}(i)=\omega_{\R}(G_{x_i})+\omega_{\R}(F_a)+\omega_{\R}(F_b)+\omega_{\R}(F_c)$. 
  
   Case 1. Assume that $\omega_{\R}(F_a)\ge \omega_{\R'}(F_a)$, $\omega_{\R}(F_b)\ge \omega_{\R'}(F_b)$, $\omega_{\R}(F_c)\ge \omega_{\R'}(F_c)$. Then $\omega_{\R}(G_{x_i})+\omega_{\R}(F_a)+\omega_{\R}(F_b)+\omega_{\R}(F_c) \ge \omega_{\R'}(G_{x_i})+\omega_{\R'}(F_a)+\omega_{\R'}(F_b)+\omega_{\R'}(F_c)$, {\em i.e.} $\Omega_{\R}(i) \ge \Omega_{\R'}(i)$. 
   
   Case 2. Assume that $\omega_{\R}(F_a)=4$, $\omega_{\R'}(F_a)=5$, $\omega_{\R}(F_b)\ge \omega_{\R'}(F_b)$, $\omega_{\R}(F_c)\ge \omega_{\R'}(F_c)$. Since $\omega_{\R'}(F_a)=5$, we have that $x^1_i$ is under the border line (in $\R'$). Because of the transformation rules, at the beginning of the proof, we have that $x^1_i$, $x^2_i$ are under the border line (in $\R$ and $\R'$), while $x^3_i$ is above the line (in $\R$ and $\R'$).
   
   Let $y_1$ be a literal of variable $x_s$ ({\em i.e.} $y_1\in \{x^1_s,x^2_s,x^3_s\}$) connected with $r_{a_2}\in V(F_a)$ via transfer. Since $\omega_{\R}(F_a)=4$, $\omega_{\R'}(F_a)=5$, we have that $y_1$ is above the border line in $\R$, and under the line in $\R'$, hence $y_1=x^3_s$, $\omega_{\R'}(F_{a'})=\omega_{\R'}(F_{b'})=4$, $\omega_{\R}(G_{x_s}) \ge 19$, where $F_{a'},F_{b'}$ are connected with $x^1_s,x^2_s$ via transfers. We have $\omega_{\R}(F_{a'})\ge 4=\omega_{\R'}(F_{a'})$ and $\omega_{\R}(F_{b'})\ge 4=\omega_{\R'}(F_{b'})$. 
   
   From the previous arguments, $\omega_{\R}(G_{x_s})+\omega_{\R}(F_a)+\omega_{\R}(F_a) \ge $
   $19+4+4 =$ $17+5+5 =$ $\omega_{\R'}(G_{x_s})+\omega_{\R'}(F_a)+\omega_{\R'}(F_a)$. 
   
   Finally, $\big( \omega_{\R}(G_{x_i})+\omega_{\R}(F_a)+\omega_{\R}(F_b)+\omega_{\R}(F_c) \big) + $ $\big( \omega_{\R}(G_{x_s})+\omega_{\R}(F_{a'})+\omega_{\R}(F_{b'})+\omega_{\R}(F_{a}) \big) \ge $
   $\big( \omega_{\R'}(G_{x_i})+ \omega_{\R'}(F_a) +\omega_{\R'}(F_b)+\omega_{\R'}(F_c) \big) + $ $\big( \omega_{\R'}(G_{x_s}) + \omega_{\R'}(F_{a'}) +\omega_{\R'}(F_{b'})+\omega_{\R'}(F_{a}) \big)$, i.e. $\Omega_{\R}(i)+\Omega_{\R}(s) \ge \Omega_{\R'}(i)+\Omega_{\R'}(s)$.

   Case 3. Assume that $\omega_{\R}(F_b)=4$, $\omega_{\R'}(F_b)=5$, $\omega_{\R}(F_a)\ge \omega_{\R'}(F_a)$, $\omega_{\R}(F_c)\ge \omega_{\R'}(F_c)$. This case is analogous to Case 2.
  
   Case 4. Assume that $\omega_{\R}(F_c)=4$, $\omega_{\R'}(F_c)=5$, $\omega_{\R}(F_a)\ge \omega_{\R'}(F_a)$, $\omega_{\R}(F_b)\ge \omega_{\R'}(F_b)$. Then $x^3_i$ is under, and $x^1_i,x^2_i$ are above the border line in $\R'$. We have two subcases.
   
   Case 4.1. Assume that $x^1_i$ or $x^2_i$ was on the same side of the line as $x^3_i$ (in $\R$). Then $\omega_{\R}(G_{x_i})\ge 19$. Hence $\omega_{\R}(G_{x_i})+\omega_{\R}(F_a)+\omega_{\R}(F_b)+\omega_{\R}(F_c) \ge $
   $19 + \omega_{\R'}(F_a)+\omega_{\R'}(F_b)+ 4 >$
   $17 + \omega_{\R'}(F_a)+\omega_{\R'}(F_b)+ 5 =$
   $\omega_{\R'}(G_{x_i})+\omega_{\R'}(F_a)+\omega_{\R'}(F_b)+\omega_{\R'}(F_c)$, i.e. $\Omega_{\R}(i) > \Omega_{\R'}(i)$. 
   
   Case 4.2. Assume that $x^1_i$ and $x^2_i$ were not on the same side of the line as $x^3_i$ (in $\R$). Then $x^3_i$ is under the line (in $\R$ and $\R'$). Let $y_3\in \{x^1_l,x^2_l,x^3_l\}$ and it is connected with $r_{c_2}\in V(F_c)$ via transfer. From $\omega_{\R}(F_c)=4$, $\omega_{\R'}(F_c)=5$, we have that $y_3$ in $\R$ was above the line, and in $\R'$ is under the line, hence $y_3=x^3_l$, $\omega_{\R}(G_{x_l}) \ge 19$, $\omega_{\R'}(F_{a''}) = \omega_{\R'}(F_{b''}) = 4$, where $F_{a''}$ and $F_{b''}$ are connected with $x^1_l$ and $x^2_l$ via transfers.
   
   It follows that $\omega_{\R}(G_{x_l}) + \omega_{\R}(F_c) + \omega_{\R}(F_c) \ge 19+4+4 =$
   $17+5+5=$
   $\omega_{\R'}(G_{x_l}) + \omega_{\R'}(F_c) + \omega_{\R'}(F_c)$.    
   
   Next, $\big( \omega_{\R}(G_{x_i})+\omega_{\R}(F_a)+\omega_{\R}(F_b)+\omega_{\R}(F_c) \big) + $ $\big( \omega_{\R}(G_{x_l})+\omega_{\R}(F_{   a''})+\omega_{\R}(F_{b''})+\omega_{\R}(F_{c}) \big) \ge $
   $\big( \omega_{\R'}(G_{x_i})+\omega_{\R'}(F_a)+\omega_{\R'}(F_b)+\omega_{\R'}(F_c) \big) + $ $\big( \omega_{\R'}(G_{x_l})+\omega_{\R'}(F_{a''})+\omega_{\R'}(F_{b''})+\omega_{\R'}(F_{c}) \big)$, i.e. $\Omega_{\R}(i)+\Omega_{\R}(l) \ge \Omega_{\R'}(i)+\Omega_{\R'}(l)$.

   Case 5. Assume that $\omega_{\R}(F_a)=\omega_{\R}(F_b)=4$, $\omega_{\R'}(F_a)=\omega_{\R'}(F_b)=5$, and $\omega_{\R}(F_c)\ge \omega_{\R'}(F_c)$. By a similar argument as in the previous cases, we have that $x^1_i,x^2_i$ are under the line (in $\R$ and $\R'$), while $x^3_i$ is above the line (in $\R$ and $\R'$). Let $y_1\in \{x^1_r,x^2_r,x^3_r\}$ be connected with $r_{a_2}\in V(F_a)$, and $y_2\in \{x^1_t,x^2_t,x^3_t\}$ be connected with $r_{b_2}\in V(F_b)$. As in the previous cases, we have $y_1=x^3_r$, $y_2=x^3_t$, and they were above the line in $\R$, and under the line in $\R'$. Hence $\omega_{\R}(G_{x_r}) \ge 19$ and $\omega_{\R}(G_{x_t}) \ge 19$. Let $x^1_r,x^2_r,x^1_t,x^2_t$ be connected with $F_{a_r},F_{b_r},F_{a_t},F_{b_t}$. Then $\omega_{\R'}(F_{a_r})=\omega_{\R'}(F_{b_r})=\omega_{\R'}(F_{a_t})=\omega_{\R'}(F_{b_t})=4$.      
   
   Therefore $\omega_{\R}(G_{x_r}) + \omega_{\R}(G_{x_t}) + \omega_{\R}(F_a) + \omega_{\R}(F_a) + \omega_{\R}(F_b) + \omega_{\R}(F_b) \ge$ $19+19+4+4+4+4 = 17+17+5+5+5+5 = $
  $\omega_{\R'}(G_{x_r}) + \omega_{\R'}(G_{x_t}) + \omega_{\R'}(F_a) + \omega_{\R'}(F_a) + \omega_{\R'}(F_b) + \omega_{\R'}(F_b)$.
  
  Hence $\big( \omega_{\R}(G_{x_i})+\omega_{\R}(F_a)+\omega_{\R}(F_b)+\omega_{\R}(F_c) \big) +$
  $\big( \omega_{\R}(G_{x_r})+\omega_{\R}(F_{a_r})+\omega_{\R}(F_{b_r})+\omega_{\R}(F_a) \big) +$
  $\big( \omega_{\R}(G_{x_t})+\omega_{\R}(F_{a_t})+\omega_{\R}(F_{b_t})+\omega_{\R}(F_b) \big) \ge $
  $\big( \omega_{\R'}(G_{x_i})+\omega_{\R'}(F_a)+\omega_{\R'}(F_b)+\omega_{\R'}(F_c) \big) +$
  $\big( \omega_{\R'}(G_{x_r})+\omega_{\R'}(F_{a_r})+\omega_{\R'}(F_{b_r})+\omega_{\R'}(F_a) \big) +$
  $\big( \omega_{\R'}(G_{x_t})+\omega_{\R'}(F_{a_t})+\omega_{\R'}(F_{b_t})+\omega_{\R'}(F_b) \big)$, i.e. $\Omega_{\R}(i) + \Omega_{\R}(r) + \Omega_{\R}(t) \ge$ $\Omega_{\R'}(i) + \Omega_{\R'}(r) + \Omega_{\R'}(t)$.
  
  Note that the case when $\omega_{\R}(F_a)=\omega_{\R}(F_b)=\omega_{\R}(F_c) = 4 < \omega_{\R'}(F_a)=\omega_{\R'}(F_b)=\omega_{\R'}(F_c) = 5$ is not possible.

  Every $i\in \{1,\ldots, n\}$ belongs to exactly one case. Variables $s$ (from Cases 2 and 3), $l$ (Case 4.2), $t$ and $r$ (Case 5) are equal to some $i\in \{1,\ldots, n\}$, but are different among themselves, i.e. there is no value that repeats itself among variables $s,l,r,t$.
 Let $A_1$ be the set of all values of $i$ from the Case 1 that are different from all $s,l,r,t$. In a similar manner we introduce sets $A_{2,3}$, $A_{4.1}$, $A_{4.2}$, $A_{5}$
  
  We will use the previous cases to prove $\omega(\R) \ge \omega(\R')$. We have $2\cdot\omega(\R) \ge \sum_{i}\omega_{\R}(G_{x_i}) + \sum_{i}\Omega_{\R}(i) = $  
  $\sum_{i}\omega_{\R}(G_{x_i})$ $+$
  $\sum_{A_1} \Omega_{\R}(i)$  $+$
  $\sum_{A_{2,3}}\big( \Omega_{\R}(i) +  \Omega_{\R}(s) \big)$ $+$
  $\sum_{A_{4.1}} \Omega_{\R}(i)$  $+$
  $\sum_{A_{4.2}}\big( \Omega_{\R}(i) + \Omega_{\R}(l)\big)$ $+$
  $\sum_{A_5}\big( \Omega_{\R}(i) + \Omega_{\R}(r) + \Omega_{\R}(t) \big)$
  $\ge$
  $\sum_{i}\omega_{\R'}(G_{x_i})$ $+$
  $\sum_{A_1} \Omega_{\R'}(i)$  $+$
  $\sum_{A_{2,3}}\big( \Omega_{\R'}(i) +  \Omega_{\R'}(s) \big)$ $+$
  $\sum_{A_{4.1}} \Omega_{\R'}(i)$  $+$
  $\sum_{A_{4.2}}\big( \Omega_{\R'}(i) + \Omega_{\R'}(l)\big)$ $+$
  $\sum_{A_5}\big( \Omega_{\R'}(i) + \Omega_{\R'}(r) + \Omega_{\R'}(t) \big)$
  $=$
  $2\cdot \omega(\R')$.
  
  Finally, $\omega(\R) \ge \omega(\R')$. Therefore $\R'$ is an optimal reconciliation. \qed    
  \end{proof}

\begin{theorem}\label{th:min_DTLCT_R_NP_hard}
	\textproc{Minimum $DTLCT_R$ reconciliation} problem is NP-hard.
\end{theorem}
\begin{proof}
	We will use a reduction from optimization version of \textproc{Max 2-Sat}. Let $F=C_1 \land C_2 \land \ldots \land C_m$, $C_j=x'_{j_1} \lor x'_{j_2}$, $j=1,\ldots,m$ be an instance of \textproc{Max 2-Sat}. 
	Trees $S$ and $G$ can be obtained in the polynomial time. After obtaining an optimal reconciliation between $S$ and $G$ as an output of \textproc{Minimum $DTLCT_R$ reconciliation}, we can (in polynomial time) obtain a proper optimal reconciliation (the proof of Theorem \ref{th:optimalProper}), and from it an optimal logical formula, i.e. a logical formula $F$ with minimum number of false clauses (Lemma \ref{formula->properReconciliation}).    \qed 
\end{proof}

Since a proper reconciliation is a $T_R$ reconciliation, {\em i.e.} it has only transfers with replacement and all losses are free, then the next theorem can be proved in the same manner as Theorem \ref{th:min_DTLCT_R_NP_hard}.

\begin{theorem}\label{th:minTRisNPHard}
	Let $A\subseteq \{D,T,L,C,T_R\}$ and $T_R\in A$. Then \textproc{Minimum A reconciliation} problem is NP-hard.
\end{theorem}

%% file: fpt_algorithm.tex
\section{\textproc{Minimum $\mathbf{T_R}$ Reconciliation} problem is fixed parameter tractable}
We will give a branch and bound algorithm that solves \textproc{Minimum $T_R$ Reconciliation} problem, with complexity $O(f(k)p(n))$, where $p$ is a polynomial, $k$ is a parameter representing an upper bound for the reconciliation's weight, and $f$ is a (computable) function. 

\subsection{Normalized reconciliation}

In order to reduce the search space of an FPT algorithm that searches for an optimal reconciliation, we introduce the notion of {\em normalized reconciliation}. The principle will be that our algorithm can output every normalized reconciliation with a non null probability, and on the other size every reconciliation can be transformed into a normalized one, without changing its cost, by operations that we call {\em node raising} and {\em transfer adjustment}.
Figures  \ref{fig:nodeRaising} and \ref{fig:transferAdjustment} depict node raising and transfer adjustment, which are defined as follows.

\begin{definition}[The reduction of an extension of a gene tree]
	If $G$ is a gene tree, and $G'$ is an extension of $G$. With $r(G')$ we denote {\em the reduction} of $G'$, obtained by pruning all lost subtrees.
\end{definition}

We can obtain $G$ from $r(G')$ by suppressing nodes of degree 2.

The next definition introduces {\em transfer adjustment} (Figure \ref{fig:transferAdjustment}). The purpose of this operation is that all transfer parents are in $V(G)$.

\begin{definition}[Transfer adjustment]\label{def:transferAdjustment}
	Let $\R=(G,G',S,\phi, \rho,\tau,\delta_T,\delta_D,\{T_R\})$ be a $T_R$ reconciliation, $x'\in V(G')\backslash V(G)$, $(x',x_t)\in \Theta(\R)$, $(x',x_t)$ is a sub-branch of a branch in $G$, $x_G$ is the minimal ancestor of $x'$ in $V(G)$. Let $\R'=(G,G'',S,\phi, \rho',\tau,\delta'_T,\delta'_D,\{T_R\})$ be a TR reconciliation such that:
	\begin{enumerate}[$(a)$]
		\item if $x_G$ is a transfer parent and $(x_G,x'_t)\in \Theta(\R)$, $x'<x_1<\ldots<x_k<x_G$, then remove $(x',x_t)$, suppress $x'_t$ and $x'$, $\rho'(x_G)=\rho(x'_t)$,  $(x_G,x_t)\in \Theta(\R')$. Next, insert new nodes $x'',x''_t$ such that $\rho'(x'')=\rho(x_G)$, $x_k<_{G''}x''<_{G''}p_{G'}(x_G)$, $\rho'(x''_t)=\rho'(x_G)$, $x_G<_{G''}x''_t<_{G''}x''$. For all  $x\in V(G')\backslash \{x_G,x',x'_t\}$ we have $\rho'(x)=\rho(x)$.
		
		\item if $x_G$ is a speciation, and $x'<x_1<\ldots<x_k<x_G$, then $\rho'(x_G)=p_{S'}(\rho(x_G))$, $(x_G,x_t)\in \Theta(\R')$, $x_{k+1}\in V(G'')$, $x_1<_{G''}\ldots <_{G''} x_{k}<_{G''}x_{x+1}$, $x_{x+1}$ is a child of $x_G$ (in $G''$), and $\rho'(x_{k+1})=\rho(x_G)$. Next, $(x',x_t)$ is removed, $x'$ is suppressed, and $\rho'(x)=\rho(x)$, $\forall x\in V(G')\backslash\{x',x_G\}$. 
	\end{enumerate}
	We say that the transfer $(x',x_t)\in\Theta(\R)$ is {\em adjusted}.
\end{definition}

The next definition introduces {\em node raising} (Figure \ref{fig:nodeRaising}). The purpose of this operation is transforming a reconciliation in a more suitable form.

\begin{definition}[Node raising]
Let $\R=(G,G',S,\phi, \rho,\tau,\delta_T,\delta_D,\{T_R\})$ be an optimal $T_R$ reconciliation, $x'\in V(G')\backslash (\Sigma(\R)\cup L(G'))$, $\rho(x')=y_1\in E(S')$, $y_1<y_2\in E(S')$. If there exists a $T_R$ reconciliation $\R'=(G,G'',S,\phi,\rho',\tau,$ $\delta'_T,\delta'_D,\{T_R\})$ such that $r(G')$ can be obtained from $r(G'')$ by supressing some nodes of degree 2, $\rho'(x')=y_2$,  and $\rho'(x)=\rho(x)$ ($\forall x\in V(r(G'))\backslash \{x'\}$),
then we say that $\R'$ is obtained from $\R$ by {\em raising node} $x'$.
\end{definition}

Note that we do not raise speciations, and do not place raised nodes at speciations from $S$.

Note that node raising and transfer adjustment does not change the number of transfers.

The next definition introduces the notion of {\em normalized reconciliation}, which represents the output of the main algorithm. The purpose of introducing this type of reconciliation is to reduce the search space of the branch and bound algorithm that we are going to give.

\begin{definition}[Normalized reconciliation]\label{def:normalizedReconciliation}
	Let $\R=(G,G',S,\phi,\rho,\tau,\delta_T,$ $\delta_D,\{T_R\})$ be an optimal $T_R$ reconciliation, and for every transfer $(x',x_t)\in E(G')$ we have $x'\in V(G)$. Let $y\in V(G')$ be the maximal element such that $x'\le y$,  $\rho(x')\le \rho(y)$, and $\tau(y) \le \tau(x')+1$. Then ($\tau(y)=\tau(x')$ and $deg(y)=2$) or ($\tau(y)=\tau(x')+1$ and $y\in \Sigma(\R)\cap V(G)$) or ($\tau(y)=\tau(x')+1$ and $y=root(G)$). If $(x',x_t)$ is a diagonal transfer, $l$ is a loss assigned to $x_t$, and $T_l$ is a lost subtree with a leaf $l$, then $|E(T_l)|=1$. Reconciliation $\R$ is called a {\em normalized reconciliation}. 
\end{definition}

The proof of the next theorem describes how to construct a normalized reconciliation from an arbitrary optimal reconciliation.
\begin{theorem}\label{th:optimalNormalized}
	Let $\R$ be an optimal reconciliation, then there exists a normalized reconciliation $\R'$ such that $\omega(\R')=\omega(\R)$.
\end{theorem}

\begin{proof}

  First, adjust all transfers (Definition \ref{def:transferAdjustment} and Figure \ref{fig:transferAdjustment}). Transfer adjustment does not change the number of transfers. Therefore the weight of reconciliation is not changed.

  We will describe how to raise transfer nodes in order to obtain a normalized reconciliation. From the transfer adjustments we have that if $(x', x_t)$ is a transfer, then $x'\in V(G)$. We can have three cases. 
  
  Case 1. There is $y\in V(G')$ such that $x'<y$, there is no transfer in the $y-x'$ path in $G'$ (i.e. $\rho(x')\le \rho(y)$ and there is no node of degree 2 in the $y-x'$ path), and $y\in \Sigma(\R)\cap V(G)$. 
   
  Case 2. There is  $y\in V(G')$ such that $x'<y$, there is neither transfer nor speciation from $V(G)$ in the $y-x'$ path in $G'$, and $deg(y)=2$. Then $y$ is a transfer child. Let $l$ be a loss assigned to $y$ and $T_l$ a lost subtree with a leaf $l$. 
  
  Case 3. There is no $y\in V(G')$ that satisfies Case 1 or 2. In this case, there is no transfer and no speciation from $V(G)$ in the path (inside $G'$) from $root(G)$ to $x'$.   
  
  First, raise all $x'\in V(G)$ that satisfy Case 1 to obtain $\tau(x')=\tau(y)-1$, where $y$ is the minimal node from Case 1.

  Then, raise all transfer children $y$ to obtain $\tau(y)=\tau(p_{G'}(y))$ or $|E(T_l)|=1$, where $T_l$ is a lost subtree with a leaf (i.e. loss) assigned to $y$.
  
  Next, raise $x'$ (from Case 2) to obtain $\tau(x')=\tau(y)$.
 
  If $x'$ satisfies Case 3, then raise it to obtain $\rho(x')=root_E(G)$.  
    
  In this way we obtain a reconciliation $\R'$. The previous procedure does not move speciations from $V(G')$. Therefore the number of transfers is not increased. Since $\R$ is an optimal reconciliation, we cannot decrease the number of transfers. Hence $\omega(\R')=\omega(\R)$. 
  
  Let us prove that $\R'$ is a normalized reconciliation. Let $(x', x_t)$ be a transfer. From the transfer adjustments, we have $x'\in V(G)$.  Let $y\in V(G')$ be the maximal element such that $x'\le y$, $\rho(x') \le \rho(y)$, and $\tau(y) \le \tau(x')+1$. Observe two cases.
  
  Case $(a)$. Assume that $\tau(y)=\tau(x')$. We need to prove that $y$ is a transfer child, {\em i.e.} $deg(y)=2$. Assume the opposite, $y$ is not a transfer child. Since $\tau(y)=\tau(x')$, $y$ is not a speciation. Therefore $y$ is a transfer parent. Transfer parents are raised as described in Cases 1, 2, and 3. Hence there is $y'$, such that $\tau(y')=\tau(y)=\tau(x)$, or $\tau(y')=\tau(y)+1=\tau(x)+1$, and $y'$ is a transfer child, speciation from $V(G)$, or $root(G)$. Since $y$ is a transfer parent, we have $y\ne y'$, {\em i.e.} $y<y'$, which contradicts the maximality of $y$. We get that $y$ is a transfer child.    
  
  Case $(b)$. Assume that $\tau(y)=\tau(x')+1$. We need to prove that $y\in V(G)\cap \Sigma$, or $y=root(G)$. Since $x'$ is a transfer parent, we have $\rho(x')\in E(S')$. From this and $\tau(y)=\tau(x')+1$, we have $\rho(y)\in V(S)$ (see assumptions about dating $S'$ and Figure \ref{fig:subdivision}). Therefore $y\in \Sigma(\R')$ or $y=root(G)$. If $y=root(G)$, this case is finished. If $y\in \Sigma(\R')$, we need to prove $y\in V(G)$. If $y\notin V(G)$, then this contradicts Cases 1, 2, 3 and the procedure of node raising.

  If $(x', x_t)$ is a diagonal transfer, then $\tau(x_t)<\tau(x')$. By the procedure for node raising, we have $|E(T_l)|=1$, where $l$ is a loss assigned to $x_t$ and $T_l$ is the lost subtree with a leaf $l$.
  
  All conditions of Definition \ref{def:normalizedReconciliation} are satisfied. Thus $\R'$ is a normalized reconciliation.  \qed   
\end{proof}

\subsection{Random normalized optimal reconciliation}\label{sec:FPT_description}

In this subsection we describe an FPT algorithm that returns a normalized reconciliation with the weight not greater than $k$, if there is one. 

The problem definition follows.

\noindent\textproc{Parametric Optimal R Reconciliation}\\
\textbf{Input:} $G,S,k\ge 0$;\\ 
\textbf{Output:} Is there an optimal reconciliation $\R=\R(G,S)$ such that $\omega(\R)\le k$? If yes, return one such reconciliation.

We are given $S$, $G$ and $\phi$, which is in this particular case a bijection between the leaves of $G$ and $S$ (Figure \ref{fig:BB_1} $(i)$). Let $A_i$ be the extant species (leafs of $S$), and $a_i$ are the extant genes (leafs of $G$)  $(i=1,\ldots, n)$. We will maintain during the execution of the algorithm a set of {\em active edges} which initially contains the terminal edges of $G$, {\em i.e.} the edges with a leaf as an extremity. Every active edge {\em belongs} to an edge in $S$, which initially is the edge determined by $\phi$. Some of the active edges might be {\em lost}, while initially none is.

Observe one time slice. Let $s_0$ be the internal node of $S$ in this time slice. Let $E_1, E_2$ be edges from $S$ incident with $s_0$ and $e_1, e_2$ active edges that belong to $E_1, E_2$ (Figure \ref{fig:BB_1} $(b)$). We have several cases.

Case 1. At least one of the edges $e_1$ or $e_2$ is lost. Then coalesce them at $s_0$ (meaning the lca of the two edges in $G$ is mapped to $s_0$ by $\rho$), and the edge that is not lost propagates to the next time slice, as well as all other edges (Figure \ref{fig:BB_1} $(c)$), where they remain active.    

Case 2. Edges $e_1$ and $e_2$ are incident. Then coalesce them at $s_0$ (Figure \ref{fig:BB_1} $(d)$).  All other active edges propagate to the next time slice, where they remain active. The parent edge of $e_1$ and $e_2$ is  also an active edge in the next time slice. 

Case 3. Edges $e_1$ and $e_2$ are neither lost nor incident.  Branch and bound tree is branching into three subtrees (subcases $(a_1)$, $(a_2)$, and $(b)$).  

Case 3--$(a_1)$. Put $e_1$ {\em on hold} (Figure \ref{fig:BB_2} $(a)$). This means that $e_1$ is not propagated into the next time slice, but stays active as long as it does not become a (diagonal) transfer (see Case 3--$(b)$). Edge $e_2$ and all other active edges from the current time slice are propagated into the next time slice. 

Case 3--$(a_2)$. The same as Case 3--$(a_1)$, but $e_2$ is {\em on hold} instead of $e_1$.

Case 3--$(b)$. Let $x$ be the minimum node in $V(G)$ that is an ancestor of both $e_1$ and $e_2$, $x'_1,\ldots, x'_{k_1}$ are the vertices in the path from $e_1$ to $x$, and $x''_1,\ldots, x''_{k_2}$ are the vertices in the path from $e_2$ to $x$ (Figure \ref{fig:BB_2} $(b)$). Take $\rho(x)=s_0$. Observe $x'_1$. Let $e_3$ be an active edge that is a descendant of $x'_1$, $E_3$ is the edge from $E(S)$ that contains $e_3$, and $x^1_1,\ldots, x^{m1}_1$ are the vertices in the path from $e_3$ to $x'_1$. Add these vertices and corresponding edges to $E_3$, as well as transfer $(x'_1, x^1_1)$. Repeat the process for every child of $x'_1$. It is possible that some of the added transfers is diagonal. We say that $x'_1$ is {\em expanded}. In $E_3$ add a lost edge $e'_3$, that is propagated to the next time slice, as an active edge. Next, expand the remaining nodes $x'_2,\ldots, x'_{k_1}$, $x''_1,\ldots, x''_{k_2}$.         

Case 4. We reached $root_E(S)$. Then expand all the remaining nodes from $V(G)$ and $\rho(root(G))=root(S)$. 

The rest of the procedure is standard branch and bound. When we reach the first solution (reconciliation) with at most $k$ transfers, we denote it by $\R^*$. If $\R$ is some other reconciliation, obtained in the branch and bound process, such that $\omega(\R)<\omega(\R^*)$, then we take $\R^*=\R$. If $\omega(\R)=\omega(\R^*)$, then we randomly take $\R^*=\R$.

If during the branch and bound procedure, we obtain a (partial) reconciliation with more than $k$ transfers, then we do not branch, and go one step back.

\subsection{Pseudocode and properties} 

In this section we give pseudocodes, prove some properties of the algorithm, and give a proof that \textproc{Minimum $T_R$ Reconciliation} is fixed parameter tractable.

\begin{algorithm}[h]
	\caption{Parametric optimal reconciliation}\label{alg:ParametricOptimalReconciliation}
	\begin{algorithmic}[1]
		\Procedure{ParametricOptimalReconciliation}{$G, S, k$} 
		  \State create $S'$ - a subdivision of $S$
		  \State $\R$ denotes partially constructed reconciliation
		  \State $\R^*$ denotes current optimal reconciliation
		  \State \textproc{Initialize}$(\R, \R^*, curr\_time\_slice)$
		  \State \textproc{BranchAndBound}$(\R, \R^*, curr\_time\_slice, k)$
		  \State \textbf{return}$(\R)$
		\EndProcedure
	\end{algorithmic}	
\end{algorithm}

\begin{algorithm}[h]
	\caption{Initializes parameters.}\label{alg:Initialize}
	\begin{algorithmic}[1]
	\Procedure{Initialize}{$\R, \R^*, curr\_time\_slice$} 
	  \State $\R^* \leftarrow NULL$
	  \State $\omega(NULL) \leftarrow +\infty$
	  \State $curr\_time\_slice \leftarrow 1$ 
	  \State assign extant genes $a_i$ to the corresponding edges $A_i$ $(i=1,\ldots, n)$

	\EndProcedure
	\end{algorithmic}	
\end{algorithm}

\begin{algorithm}[h]
	\caption{Branch and bound}\label{alg:BranchAndBound}
	\begin{algorithmic}[1]
	\Procedure{BranchAndBound}{$\R, \R^*, curr\_time\_slice, k$} 
	  \If{$root_E(S)$ is in $curr\_time\_slice$}
	     \State expand remaining nodes from $V(G)$
	     \If{$\omega(\R)<\omega(\R^*)$}
	        \State $\R^* \leftarrow \R$
	     \EndIf
	     \If{$\omega(\R) == \omega(\R^*)$}
	        \State $\R^* \leftarrow \R$ - \textbf{random}
	     \EndIf
	     \State \textbf{return} 
	  \EndIf 
	  
	  \State $state_1$ - the state of reconciliation $\R$  
      \State $s_0\in V(S)$ - speciation in $curr\_time\_slice$
      \State $E_1,E_2$ - edges of $S$ incident with $s_0$
      \State $e_1,e_2$ - active edges of $G'$ that are inside $E_1,E_2$
    
      \If{($e_1$ or $e_2$ is a lost edge) \textbf{or} ($e_1$ and $e_2$ are incident in $G$)}
        \State coalesce $e_1,e_2$ into a speciation at $s_0$
        \State all other active edges propagate to the next time slice
        \State $curr\_time\_slice++$
        \State \textproc{BranchAndBound}$(\R, \R^*, curr\_time\_slice, k)$
      \Else  
        \State \textproc{ExpandOneEdge}$(\R, \R^*, curr\_time\_slice, (a_1), k)$ 
        \State \textproc{ExpandOneEdge}$(\R, \R^*, curr\_time\_slice, (a_2), k)$      
        \State \textproc{ExpandOneEdge}$(\R, \R^*, curr\_time\_slice, (b), k)$
     \EndIf
      
	\EndProcedure
	\end{algorithmic}	
\end{algorithm}

\begin{algorithm}[H]
	\caption{Executes one edge incident with branching vertex of branch and bound tree }\label{alg:BranchAndBoundEdge}
	\begin{algorithmic}[1]
	\Procedure{ExpandOneEdge}{$\R, curr\_time\_slice, case, k$} 
	 
	 \If{$case == (a_1)$ \textbf{or} $case == (a_2)$}
	   \If{$case==(a_1)$}
	     \State  $e' \leftarrow e_1$
	   \Else
	     \State $e'\leftarrow e_2$ 
	   \EndIf 
	   
	   \State put $e'$ {\em on hold}
	   \State propagate all other active edges to the next time slice
	   \State $curr\_time\_slice++$ 
	   \State \textproc{BranchAndBound}$(\R, \R^*, curr\_time\_slice, k)$
	   \State reset $\R$ to $state_1$       
	 	
     \Else		
		\State $x \leftarrow lca_G(e_1,e_2)\in V(G)$
		\State $x'_1,\ldots, x'_{k_1}$ nodes from $V(G)$ in the path from $e_1$ to $x$  
		\State $x''_1,\ldots, x''_{k_2}$ nodes from $V(G)$ in the path from $e_2$ to $x$
		\State assign $x$ to $s_0$
		\State expand $x'_1,\ldots, x'_{k_1}$ and $x''_1,\ldots, x''_{k_2}$
	    \State $t \leftarrow \omega(\R)$ - i.e. the number of transfers in the current partial reconciliation;
		\If{$t>k$} 
	      \State \textbf{return}
		\EndIf 
\algstore{myStore}
\end{algorithmic}
\end{algorithm}

\begin{algorithm}
 \begin{algorithmic}	
   \algrestore{myStore}	
		\If{$\R$ is a (complete) reconciliation}
		  \If{$\omega(\R)<\omega(\R^*)$}
		     \State $\R^*\leftarrow \R$  \label{line:newOptimal}
		  \ElsIf {$\omega(\R) == \omega(\R^*)$}
		      \State $\R^*\leftarrow \R$ - \textbf{random}  \label{line:randomSelection}
		  \EndIf
		  
		  \State \textbf{return}
		\EndIf  
	    
	    \State $curr\_time\_slice++$
		\State \textproc{BranchAndBound}$(\R, \R^*, curr\_time\_slice)$       
		\State reset $\R$ to $state_1$
	\EndIf	
	
	\EndProcedure
	\end{algorithmic}	
\end{algorithm}

\begin{theorem}\label{th:possibleOutput}
	Let $\R$ be a normalized reconciliation and $\omega(\R)\le k$. Then $\R$ is a possible output of  Algorithm \ref{alg:ParametricOptimalReconciliation}.
\end{theorem}
\begin{proof}
	 Since $\R$ is a normalized reconciliation, it is also, by definition, optimal. If $I$ is a time slice, then $R_I$ denotes partial reconciliation induced by $I$, {\em i.e.} the part of $\R$ that is inside $I$, and all other time slices before $I$. We will prove that the algorithm constructs $\R_I$ during the execution. We will use mathematical induction on $I$.  
	
	Let $I_0$ be the first time slice, and $s_0\in V(S)$ is a speciation such that $\tau(s_0)\in I_0$ (Figure \ref{fig:BB_1}), $E_1, E_2\in E(S)$ are incident with $s_0$, $e_1,e_2$ are active edges in $E_1$ and $E_2$. 
	
    Let us prove that $\R_{I_0}$ can be obtained during the execution of the algorithm.   We have several cases.

    Case 1. Edges $e_1$ and $e_2$ are incident. Let us prove that $e_1$ and $e_2$ coalesce at $s_0$. Assume the opposite, $\rho(x)\ne s_0$, where $x\in V(G)$ is incident with both $e_1$ and $e_2$. Then $e_1$ or $e_2$ is a transfer. By placing $\rho(x)=s_0$ we decrease the number of transfers in $\R$, which contradicts the optimality of $\R$.

    Case 2. 
    Edges $e_1$ and $e_2$ are not incident. We will investigate subcases. Some subcases are not obtainable by the algorithm. For them, we will prove they cannot occur in $\R$. Let $x$ be the minimal element from $V(G)$ that is an ancestor of $e_1$ and $e_2$.
    
    Case 2.1. Let $\rho(x)=s_0$, $\rho(x'_{i_1})=E_1$, $\rho(x''_{i_2})=E_2$ $(i_1=1,\ldots,k_1, i_2=1,\ldots,k_2.)$. This case is obtainable by the algorithm.
    
    Case 2.2. Let $e_i$ contains a diagonal transfer and $e_{3-i}$ is propagated to the next time slice ($i=1$ or $i=2$). This case is also obtainable by the algorithm. 
    
    Case 2.3. Both $e_1$ and $e_2$ are propagated to the next time slice. Then $s_0$ contains two gene lineages, which is impossible for a TR reconciliation. Therefore this case cannot occur.  
    
    Case 2.4 We have $\rho(x)=s_0$, and there is  $y_1\in \{x'_1,\ldots,x'_{k_1}\}$, or $y_2\in \{x''_1,\ldots,x''_{k_2}\}$ such that $\rho(y_1)\ne E_1$, or $\rho(y_2)\ne E_2$. Since $s_0$ is the only speciation in $S$ in the current time slice, then all $x'_1,\ldots,x'_{k_1}$ and $x''_1,\ldots,x''_{k_2}$ are transfers. Therefore by moving $y_1$ (or $y_2$) into $e_1$ (or $e_2$) we remove some of the transfers. In this way, we obtain a reconciliation cheaper than $\R$, which contradicts the optimality of $\R$. Hence this case is impossible.       
	
	Case 2.5. Assume that $\tau(x) > \tau(s_0)$, $\tau(y_1)\le \tau(s_0)$, and $\tau(y_2)\le \tau(s_0)$ for some $y_1\in \{x'_1,\ldots,x'_{k_1}\}$, $y_2\in \{x''_1,\ldots,x''_{k_2}\}$. Then $\R$ is not a normalized reconciliation. Hence this case is not possible.

	Case 2.6. If $x$ is in $I_0$ and $\rho(x) \ne s_0$, then by taking $\rho(x)=s_0$ we get a reconciliation with fewer transfers, contrary to the optimality of $\R$.  

	For the inductive hypothesis part, assume that the statement is true for time slices $I_0, I_1,\ldots, I_{k-1}$. Let us prove that it is true for $I_k$. Proving the statement for $I_k$ is the same as for $I_0$, therefore we will not repeat it. 

    Hence $\R_I$ is obtainable by the procedure. Since $\R_I=\R$ for the final time slice $I$, $\R$ is also obtainable by the algorithm. Since it is an optimal reconciliation, $\R$ is a possible output of the algorithm. \qed     
\end{proof}

\begin{theorem}\label{th:optimalOutput}
	If Algorithm \ref{alg:ParametricOptimalReconciliation} returns a reconciliation $\R$, then $\omega(\R)\le k$ and $\R$ is a normalized reconciliation.
\end{theorem}
\begin{proof}
	It is obvious that $\omega(\R)\le k$, because the Algorithm cuts an edge of the branch-and-bound tree if $t>k$. 
	
	Let $(x', x_t)\in E(G')$ be a transfer, $y\in V(G')$ is the maximal element such that $x'\le y$, $\rho(x')\le \rho(y)$, $\tau(y)\le \tau(x')+1$. In the algorithm, transfers are created when nodes are expanded. Since only nodes in $V(G)$ are expanded, every transfer starts in a node from $V(G)$. Hence $x'\in V(G)$. 
	
	Transfers are constructed in the Case 3--$(b)$ and Case 4 (see Subsection \ref{sec:FPT_description}). If $x'\in \{x'_1,\ldots, x'_{k_1}, x''_1,\ldots, x''_{k_2}\}$, then $y\in V(G)\cap \Sigma(\R)$. If $y\notin V(G)\cap \Sigma(\R)$, then $(x',x_t)$ is obtained in the expanded part, and $deg(y)=2$. 
	
	If $deg(y)=2$ and $\tau(y)<\tau(p_{G'}(y))$, then $(p_{G'}(y), y)$ is a diagonal transfer. Diagonal transfers are made by using edges from $G$ that were {\em on hold}. From Case 3--$(a)$ we have that a loss $l$, assigned to $y$, belongs to a lost subtree $T_l$ with one edge and $\tau(root(T_l))=\tau(y)+1$. 
	
	If transfer $(x', x_t)$ is expanded in $root_E(S)$, then $y=root(G)$.

	Now we will prove that $\R$ is an optimal reconciliation. Assume the opposite, $\R$ is not an optimal reconciliation. From Theorem \ref{th:optimalNormalized} there is a normalized optimal reconciliation.  Let $\R'$ be a normalized reconciliation, i.e. $\omega(\R') < \omega(\R)$. 

	 From Theorem \ref{th:possibleOutput}, $\R'$ is a possible output of Algorithm \ref{alg:ParametricOptimalReconciliation}. Since the algorithm always replaces current reconciliation, with the less expensive (if it finds one), $\R$ cannot be an output of Algorithm \ref{alg:ParametricOptimalReconciliation}, since it would be replaced by $\R'$ (or some other reconciliation), a contradiction. 
	 
	 We have thus proved that $\R$ is a normalized reconciliation, and $\omega(\R)\le k$.  \qed 
\end{proof}

\begin{theorem}\label{th:complexity}
	Time complexity of Algorithm \ref{alg:ParametricOptimalReconciliation} is $O(3^k\cdot n)$.
\end{theorem}
\begin{proof}
	Branching in the algorithm occurs if and only if we add transfers, i.e. with every branching we add at least one transfer. Therefore we can have the branch depth at most $k$. Since we branch to three cases ($a_1, a_2$, and $b$),  the size of the branch and bound tree is $O(3^k\cdot n)$. \qed
\end{proof}

\begin{theorem}\label{th:minR_FPT}
	\textproc{Minimum $T_R$ Reconciliation} problem is fixed parameter tractable with respect to the parameter that represents an upper bound for the reconciliation's weight.
\end{theorem}
\begin{proof}
	Follows directly from Theorems \ref{th:optimalNormalized}, \ref{th:possibleOutput}, \ref{th:optimalOutput}, and \ref{th:complexity}. \qed
\end{proof}

%% file: equivalence_with_dated_spr.tex
\section{Minimum dated SPR scenario is NP-hard and FPT}

Finally, we prove that a constrained version of the well known SPR distance problem, the \textproc{Minimum Dated SPR Scenario}, mentioned in \cite{Song2006}, is equivalent to the \textproc{Minimum $T_R$ Reconciliation} problem.

\begin{definition}[Dated SPR operation]
	Let $T$ be a dated, fully binary, rooted tree, $e_1=(a_2, a_1)$, $e_2=(b_2, b_1)\in E(T)$, where $a_2=p(a_1)$,  $b_2=p(b_1)$, and $\tau(a_1) < \tau(b_2)$. Delete $e_1$, suppress $a_2$, subdivide $e_2$  with node $a'_2$,  where $\tau(a_1) \le \tau(a'_2)$, connect $a_1$ and $a'_2$. Obtained tree denote by $T'$. We say that $T'$ is obtained from $T$ by a dated SPR operation.   
\end{definition} 

We will denote this SPR operation by $spr((a_2,a_1),(b_2,b_1))=a'_2$. Note that if $spr((a_2,a_1),(b_2,b_1))=a'_2$, $spr((a_2,a_1),(b_2,b_1))=a''_2$, and $\tau(a'_2)\ne \tau(a''_2)$, then these two SPR operations are different.

\begin{definition}[\textproc{Minimum Dated SPR Scenario} problem]
	Let $T$ and $T'$ be rooted, fully binary trees, where $T$ is dated and $T'$ is undated tree. Assigning dates to $V(T')$, and finding a minimum number (over all possible date assignments to $V(T')$) of SPR operations that transform $T$ into $T'$ is called {\em \textproc{Minimum Dated SPR Scenario} problem}. The number of SPR operations is called {\em the length of SPR scenario}.  
\end{definition}

Now, we introduce parametrized versions of the problems we are interested in. 

\noindent\textproc{k-Minimum $T_R$ Reconciliation:}\\
\textbf{Input:} $S, G, k$.\\
\textbf{Output:} Is there an optimal $T_R$ reconciliation $\R$ such that $\omega(\R)\le k$? 

\vspace{\baselineskip}

\noindent\textproc{k-Minimum Dated SPR Scenario:}\\
\textbf{Input:} $T$ - dated, $T'$ - undated, full  binary, rooted trees\\
\textbf{Output:} Is there an optimal dated SPR reconciliation with the length not greater than $k$?

\begin{lemma}\label{lem:reduction}
	The problem \textproc{k-Minimum Dated SPR Scenario} is (polynomially) equivalent to the problem \textproc{k-Minimum $T_R$ Reconciliation}. 
\end{lemma}
\begin{proof}
 Note that if $(a_2,a_1)\in E(G)$, then there is a path in $G'$ $(a_2, b_1, \ldots, b_k, a_1)$. The length of this path is at least 1, i.e. $k\ge 0$. Hence every edge from $G$ is a path in $G'$. Also, $(a_2, a_1)$ can contain a transfer. In this proof we assume that all transfers are adjusted (as described by Definition \ref{def:transferAdjustment} and Figure \ref{fig:transferAdjustment}), i.e. all transfers start in $V(G)$.
 
  We introduce coloring of edges and nodes that were involved in some SPR operation.  Let $spr((a_2, a_1), (b_2, b_1)) = a'_2$ be the $i$-th SPR operation $T_i \rightarrow T_{i+1}$. Then we color edge $(a'_2, a_1)$ and node $a'_2$ with color $C_i$. If edge $(b_2,b_1)$ was colored, then edges $(b_2,a'_2)$ and $(a'_2,b_1)$ are colored with the same color. Let $c_1$ be the child of $a_2$ (in $T_i$) different from $a_1$, and $c_2$ is the parent of $a_2$ (in $T_i$). Then $c_2$ is the parent of $c_1$ (in $T_{i+1}$). If edge $(c_2,a_2)$ was colored with a color, then edge $(c_2,c_1)$ is colored with the same color.       
  
  To the optimal SPR scenario we will assign an optimal $T_R$ reconciliation. Colored edges will represent transfers, colored nodes will be transfer parents, non-colored edges will coincide with the edges of species tree, and non-colored nodes will be speciations.   
   
 Let us first demonstrate the reduction from \textproc{k-Minimum $T_R$ Reconciliation} to \textproc{k-Minimum Dated SPR Scenario}. 
 Let $S$ and $G$ be a species and gene tree,  $S=T_0\rightarrow T_1\rightarrow \ldots \rightarrow T_k=G$ be an optimal SPR scenario transforming $S$ into $G$. Using this optimal SPR scenario, we will construct an optimal $T_R$ reconciliation $\R=(G,G',S,\phi,\rho,\tau,\delta_T,\delta_D,\{T_R\})$.
 
 Note that in $T_k$ we have at most $k$ nodes that are colored. Also, colored edges form (colored) subtrees of $T_k$ with colored roots and inner nodes, while the leafs of these trees are not colored.   
 
 If $a\in V(T_k)$ is a non-colored node, then it  can be observed as a node from $S$ and node from $G$. Take $\rho(a)=a\in V(S)$, for all non-colored nodes $a\in V(T_k)=V(G)$. Non-colored paths connect non-colored nodes. All non-colored edges from $T_k=G$ place inside $S$ so that they contain no transfer. Note that leafs of $T_k$ are non-colored.
 
 Now, inside $S$ we will place colored nodes and colored edges.  Let $T_c$ be an arbitrary colored tree, and $c_0$ is its root. Then $c_0$ is on a non-colored path of $G$, and we will leave it there in $S$. Next, we will move inner nodes of $T_c$ so we place them inside $S$. Let $L(T_c)=\{l_1,\ldots, l_s\}$, and $\tau(l_1)\ge \ldots \ge \tau(l_s)$. Assume that $c^1_1, c^1_2, \ldots, c^1_{i_1}$ are inner nodes of $T_c$ in the path from $l_1$ to $c_0$ whose placement inside $S$ is not defined. Then place these nodes in the edge of $S'$ just above $l_1$, i.e. $\rho(c^1_1) = \ldots =\rho(c^1_{i_1})=p_{S'}(\rho(l_1))$. Repeat the previous process for leafs $l_2, \ldots l_s$. In this way we obtain a reconciliation with transfers, and every edge of S at any moment contains at most one lineage from $G'$, hence if we extend losses we obtain a $T_R$ reconciliation. Since a transfer can start only at a colored node, we have at most $k$ transfers, i.e. $\omega(\R) \le k$.  
 
 After the next reduction, we will prove that $\R$ is an optimal reconciliation. 
 
 In the second part, we  demonstrate a reduction from \textproc{k-Minimum Dated SPR Scenario} to \textproc{k-Minimum $T_R$ Reconciliation}.
  Let $T$ be a dated and $T'$ is an undated binary rooted tree. We need a minimum dated SPR scenario $T=T_0\rightarrow T_1\rightarrow \ldots \rightarrow T_k=T'$.
  
  Take $S=T$ and $G=T'$. Let $\R$ be an optimal $T_R$ reconciliation, and $\omega(\R)=k$. We will prove that the length of minimum dated SPR scenario is $k$, and reconstruct it using $\R$.
  
  First, let us construct a scenario of the length $k$. Adjust all transfers in $\R$, so they start at the nodes from $V(G)$, just like in the first step of the proof of Theorem \ref{th:optimalNormalized} (Definition \ref{def:transferAdjustment}, Figure \ref{fig:transferAdjustment}). 
  
  Take $T_k=T'$, $G_k=G$, $G'_k=G'$, and $\R_k=\R$. Let $(x_2, x_1)$ be an arbitrary transfer, $x'_1$ is the child of $x_1$ in $G'$, $l$ is the loss assigned to $x_1$, and $l_0=root(T_l)$, where $T_l$ is a lost subtree such that $l\in L(T_l)$. Let $p_k = (l_0, l_1, \ldots, l_{s-1}, l_s=l)$ be a path in $G'$ (i.e. in $T_l$), and therefore a lost path. Remove $(x_2, x_1)$ from $G'_k$, suppress $x_2$, include the path $p_k$ into $G_k$ ($p_k$ is not a lost path anymore), suppress $x_1$. Thus we eliminate one transfer, and obtain $G_{k-1}, G'_{k-1}, \R_{k-1}$, where $\omega(\R_{k-1})=\omega(\R_k)-1$. Repeating this procedure, we obtain an SPR scenario $T'=T_k\rightarrow T_{k-1}\rightarrow \ldots \rightarrow T_0=T$, i.e. $T=T_0\rightarrow T_1\rightarrow \ldots \rightarrow T_k=T'$. 
  
  Since the transfers can be horizontal or diagonal, corresponding SPR operations are dated. We proved that optimal dated SPR scenario transforming $T$ into $T'$ has the length at most $k$.  
  
  Let us prove that previous reductions construct optimal reconciliation (the first reduction) and optimal SPR scenario (the second reduction). Let $T_1 \rightarrow \ldots T_k$ be an optimal SPR scenario. Take $S=T_1, G=T_k$ and $\R$ is a reconciliation obtained in the first reduction. We have $k'=\omega(\R)\le k$. Now, let $T_1=T'_1\rightarrow T'_2\rightarrow \ldots \rightarrow T'_{k''}=T_k$ be a SPR scenario obtained from $G$ and $S$ in the second reduction. Then $k''\le k' \le k$. Since there is no SPR scenario, transforming $T_1$ into $T_k$, with the length less than $k$, we have $k''=k'=k$. \qed   
\end{proof}

\begin{theorem}\label{th:SPR_is_NP_hard}
  \textproc{Minimum Dated SPR Scenario} is NP-hard
\end{theorem}
\begin{proof}
  Since there is a polynomial reduction from \textproc{Minimum $T_R$ Reconciliation}  to \textproc{Minimum Dated SPR Scenario} (Lemma \ref{lem:reduction}) and \textproc{Minimum $T_R$ Reconciliation} is NP-hard (Theorem \ref{th:minTRisNPHard}), then  \textproc{Minimum Dated SPR Scenario} is NP-hard.    \qed              
\end{proof}

\begin{theorem}
	\textproc{Minimum Dated SPR Scenario} is FPT with respect to pa\-ram\-e\-trized distance.
\end{theorem}
\begin{proof}
	Since there is a polynomial reduction (which is also an FPT reduction) from \textproc{k-Minimum Dated SPR Scenario} to \textproc{k-Minimum $T_R$ Reconciliation} (Lemma \ref{lem:reduction}) and \textproc{Minimum $T_R$ Reconciliation} is FPT (Theorem \ref{th:minR_FPT}), then \textproc{Minimum Dated SPR Scenario} is FPT. \qed
\end{proof}

%% file: conclusion.tex
\section{Conclusion}

We propose an integration of two ways of detecting lateral gene transfers, and more generally to construct gene histories and handle the species tree gene tree discrepancies. On one side, SPR scenarios model transfers with replacements and are limited by computational complexity issues, the difficulty to include time constraints and other gene scale events like transfers without replacement, duplications, conversions and losses. On the other side, reconciliation algorithms usually work with dynamic programming, necessitating an independence hypothesis on different gene tree lineages, incompatible with replacing transfers.

We think this is a big issue for biological models, because the results can depend on the type of methodology which is chosen, leading to simplification hypotheses. Moreover, algorithms are often tested with simulations containing the same hypotheses as the inference models. This is why it can be important to explore methodological issues at the edge of both methods, which is what we do here.

Future work include imagining a way to include transfer with replacement in standard reconciliation software. This will require more integration and probably more efficient algorithms so that it does not harm the computing time.

%% file: figures.tex
\section{Figures}

\begin{figure}[H]
	\centering
	\includegraphics{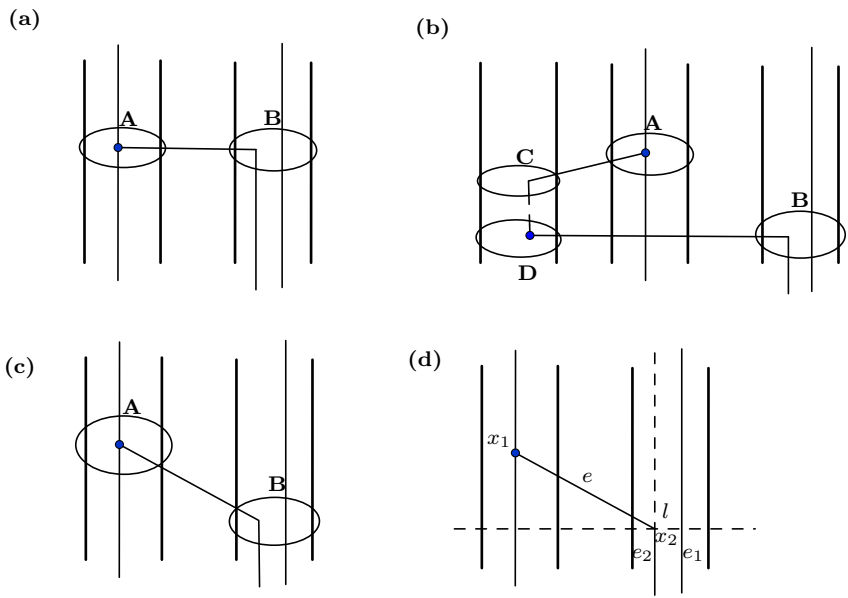}
	\caption{Gene transfer. \textbf{(a)} Horizontal gene transfer between species existing at the same moment. Species $A$ is a donor species, while species $B$ is a recipient species, and it receives a new gene copy. \textbf{(b)} Gene exits the observed phylogeny, through a speciation or transfer, then it returns through a horizontal transfer. \textbf{(c)} The event from (b) can be represented with a diagonal transfer. \textbf{(d)} Transfered gene ($x_2$) replaces already present gene ($l$). Replaced gene $l$ is lost. This event is called {\em replacement transfer} or {\em transfer with replacement}, and is represented by a transfer and gene loss. Formally, $\delta_T(e)=l$, where $e=(x_1,x_2)$ is a transfer.} 
	\label{fig:transfer}	
\end{figure}

\begin{figure}
	\centering
	\includegraphics{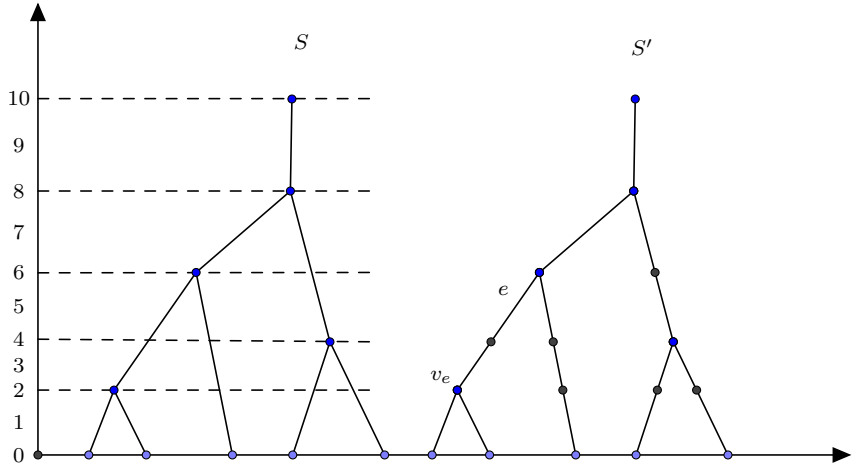}	
	\caption{Tree $S'$ denotes the subdivision of a tree $S$. To nodes from $S'$ even dates are assigned, while edges from $S'$ are assigned odd dates. The dates are integers from $0$ to $2n$, where $n$ is the number of the extant species.
	If $e\in E(S')$, then $v_e$ is the maximum node from $S$ such that $v_e < e$.}
	\label{fig:subdivision}
\end{figure}

\begin{figure}
	\centering
	\includegraphics{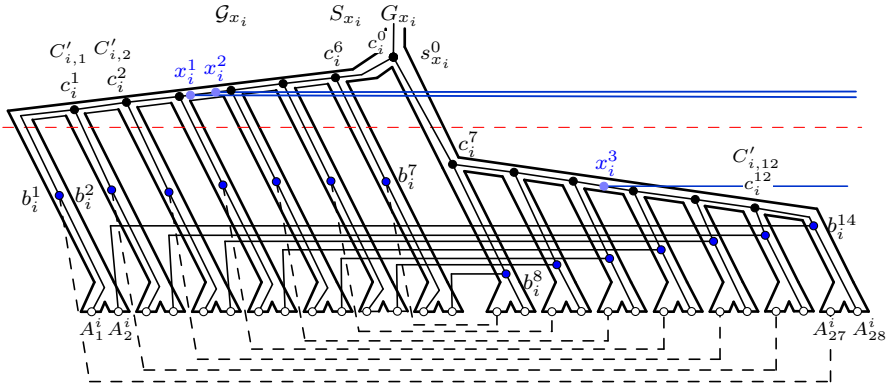}
	\caption{A variable gadget denoted by $\G_{x_i}$. It is composed of $S_{x_i}$ (a part of the species tree $S$) and $G_{x_i}$ (a part of gene tree $G$). Nodes $A^i_1,\ldots, A^i_{28}$ are leaves of $S_{x_i}$. Nodes $C'_{i,1}, \ldots, C'_{i,12}$ are some of the inner nodes of $S_{x_i}$, and $s^0_{x_i}$ is the root of $S_{x_i}$. The rest of the labels denote some of the nodes of $G_{x_i}$.  
	Variable $x_i$ has two positive (represented by $x^1_i, x^2_i\in V(G)$) and one negative literal (represented by $x^3_i\in V(G)$). We can assume that every variable has exact three literals, and there is at least one positive and one negative literal.}
	\label{fig:variable_gadget}	
\end{figure}

\begin{figure}
	\centering
	\includegraphics{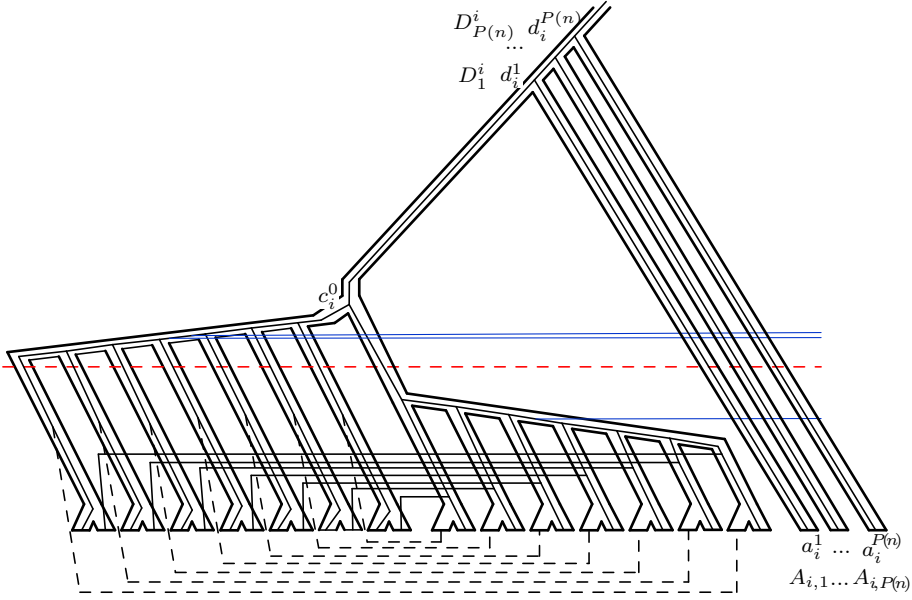}
	\caption{A variable gadget with anchor. We have $P(n)$ species in the anchor, where $P$ is sufficiently large polynomial. Nodes $d^1_i,\ldots, d^{P(n)}_i, a^1_i,\ldots, a^{P(n)}_i$ belong to gene tree that is part of anchor. }
	\label{fig:variable_gadgetb}	
\end{figure}

\begin{subfigures}
\begin{figure}
	\centering
	\includegraphics{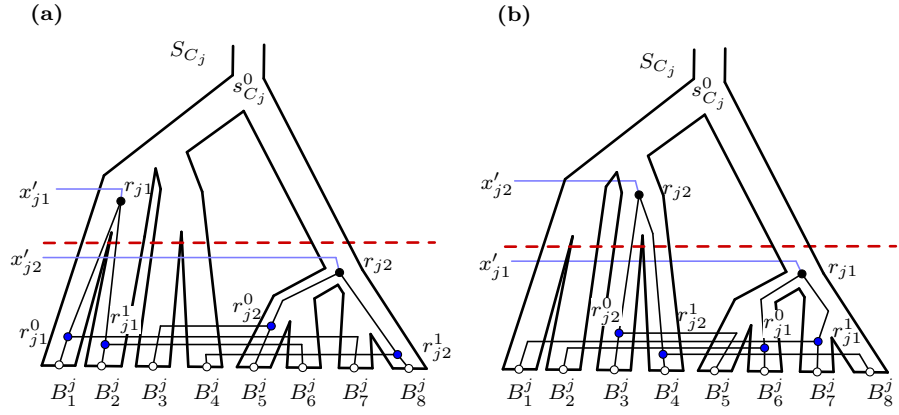}
	\caption{Clause gadget that corresponds to a clause $C_j=x'_{j_1} \lor x'_{j_2}$. \textbf{(a)} Literal $x'_{j_1}$ is true, and $x'_{j_2}$ is false. \textbf{(b)} Literal $x'_{j_2}$ is true, and $x'_{j_1}$ is false.}
	\label{fig:clause_gadgeta}	
\end{figure}
\begin{figure}
	\centering
	\includegraphics{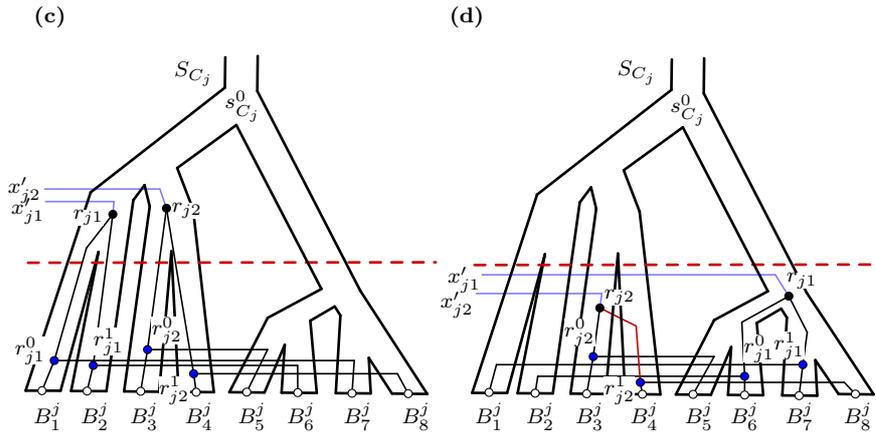}
	\caption{\textbf{(c)} Both literals are true. \textbf{(d)} Both literals are false, hence the clause is false. In this case we have an extra transfer.}
	\label{fig:clause_gadgetb}	
\end{figure}
\end{subfigures}

\begin{figure}
	\centering
	\includegraphics{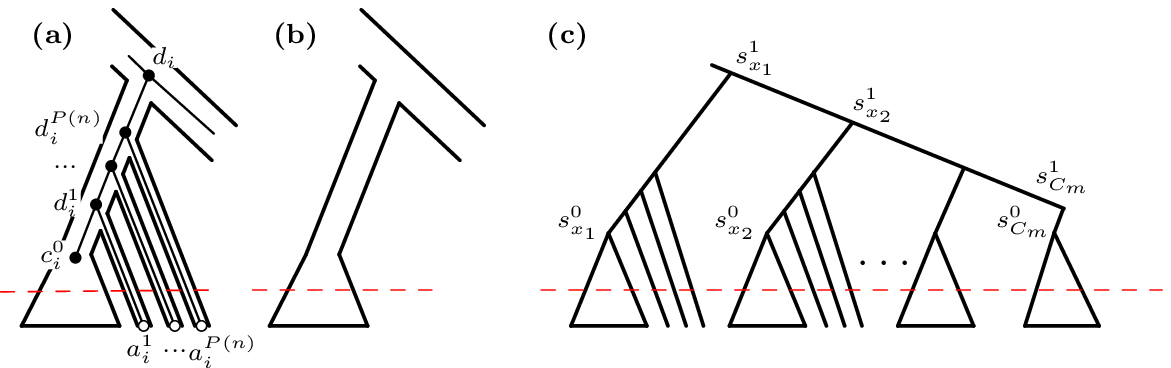}	
	\caption{\textbf{(a)} Variable gadget with anchor. For $i=n$, $d_i$ (i.e. $d_n$) does not exist. \textbf{(b)} Clause gadget. \textbf{(c)} Proper reconciliation. Nodes $s^\alpha_\beta$ ($\alpha\in\{0,1\}$, $\beta\in\{x_1,\ldots,C_m\}$) belong to species tree.}
	\label{fig:whole_reconciliation}
\end{figure}

\begin{figure}
	\centering
	\includegraphics{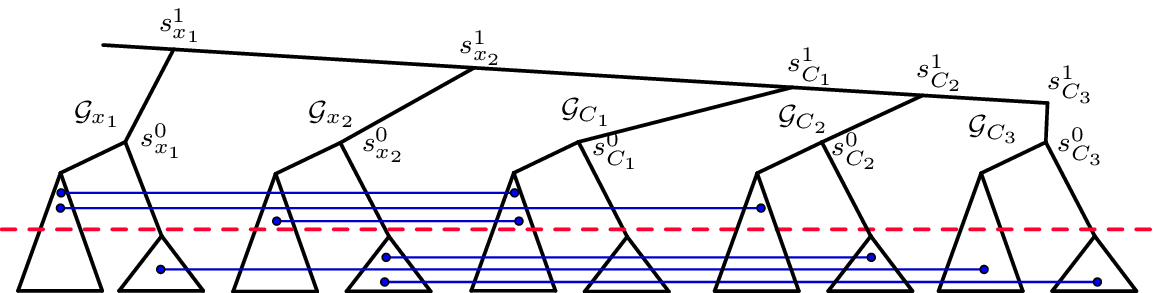}	
	\caption{A proper reconciliation assigned to formula $F_1=(x_1\lor \lnot x_2)\land(x_1\lor x_2)\land(\lnot x_1 \lor x_2)$ with values $x_1=1, x_2=0$. Some other formulas are also possible, like $F_2=(\lnot x_1\lor \lnot x_2)\land(\lnot x_1\lor x_2)\land(x_1 \lor x_2)$ with values $x_1=0, x_2=0$. Clauses $C_1$ and $C_2$ are true,  and clause $C_3$ is false.}
	\label{fig:example_whole_reconciliation}
\end{figure}

\begin{figure}
	\centering

	\includegraphics{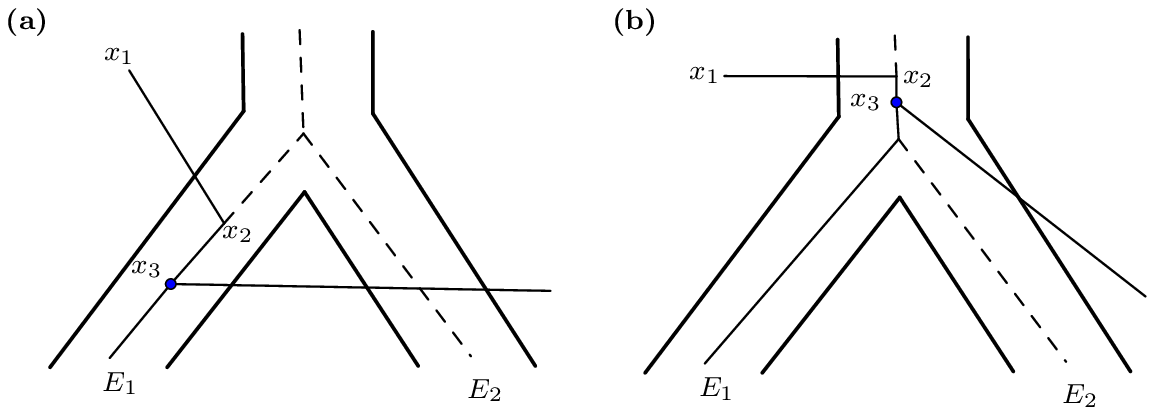}

	\caption{Node raising. \textbf{(a)} Transfer $(x_1,x_2)$, and $x_3$ is the only child of $x_2$. \textbf{(b)} We first raise $x_2$, then we raise $x_3$. Node $x_2$ cannot be raised higher than $x_1$. Node $x_3$ can be arbitrary close to $x_2$.}
	\label{fig:nodeRaising}	
\end{figure}

\begin{figure}
	\centering
	\includegraphics{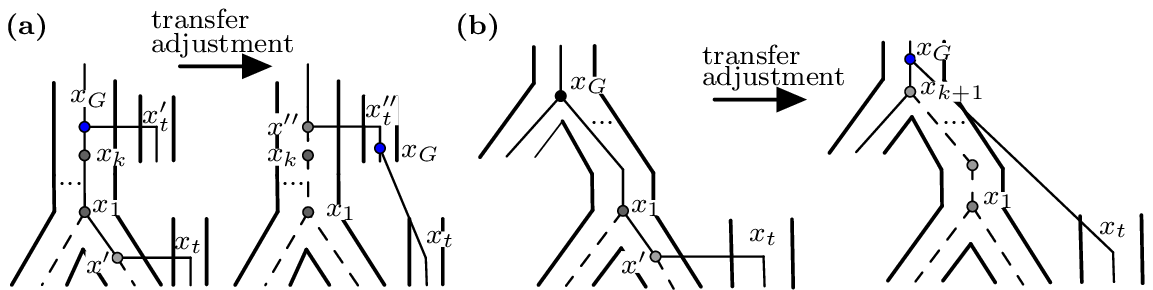}	
	\caption{Transfer adjustment. Observe transfer $(x',x_t)$, where $x'\in V(G')\backslash V(G)$. The path $(x',x_1,\ldots, x_k, x_G)$ is in $G'$, and $x_G$ is the minimum ancestor of $x'$ in $G$. We adjust transfers in order to obtain that all transfers' parents are in $V(G)$. \textbf{(a)} We have $x_G\notin \Sigma$, hence $x_G$ is a transfer parent, and $(x_G,x'_t)$ is a transfer. Node $x_G$ now denote by $x''$, and move $x_G$ to obtain $\rho(x_G)=\rho(x'_t)$ and $x'_t$ is a parent of $x_G$. New transfers are $(x'', x'_t)$ and $(x_G, x_t)$. Suppress node $x'$, $(x_1,\ldots, x_k, x'')$ is a lost path, and $x''\notin V(G)$.  \textbf{(b)} We have $x_G\in \Sigma$. Raise $x_G$, suppress $x'$, and the new transfer is $(x_G, x_t)$. The path $(x_1,\ldots,x_k,x_{k+1})$ is lost, and $x_{k+1}$ is a child of $x_G$.}
	\label{fig:transferAdjustment}
\end{figure}

\begin{figure}
	\centering
	\includegraphics{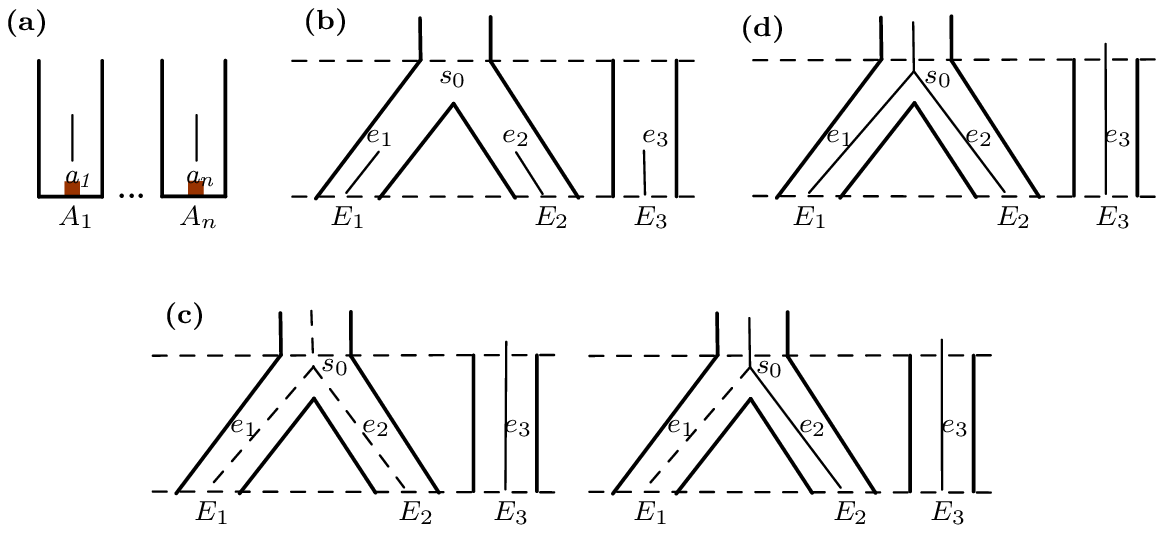}
	\caption{Initialization and Case 1. \textbf{(a)} Initializing the partial reconciliation, at the beginning of B\&B. To every extant gene is assigned an active edge. \textbf{(b)} We observe current time slice, where $s_0$ is a corresponding speciation from $S$. \textbf{(c)} If at least one of the edges $e_1$ and $e_2$ is lost, then they are coalesced at $s_0$, and non-lost edge is propagated to the next time slice, as well as all other edges from the current time slice. \textbf{(d)} If $e_1$ and $e_2$ are incident (i.e. they are siblings), then they coalesce at $s_0$.}
	\label{fig:BB_1}	
\end{figure}

\begin{figure}
	\centering
	\includegraphics{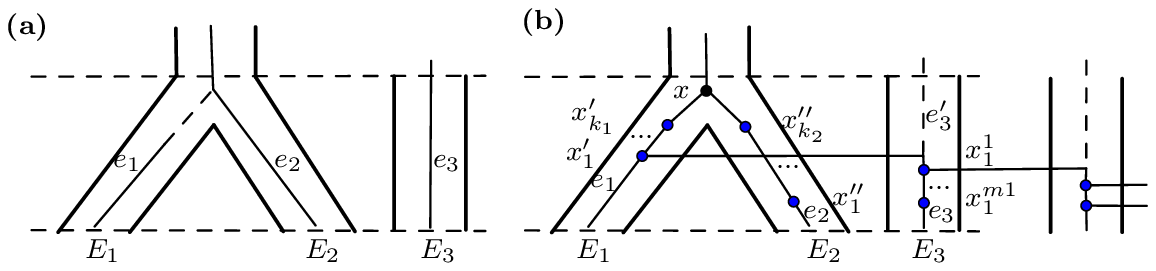}
	\caption{Cases 2 and 3. \textbf{(a)} Edge $e_1$ is put on hold (staying active), waiting to become a (diagonal) transfer. Edge $e_2$ is propagated to the next time slice, as well as all other active edges from the current time slice. \textbf{(b)} Let $x$ be the minimal ancestor of $e_1$ and $e_2$ in $V(G)$. Assign $x$ to $s_0$, and expand all nodes between $x$ and $e_1$, and between $x$ and $e_2$.}
	\label{fig:BB_2}	
\end{figure}